\documentclass[12pt, draftclsnofoot, onecolumn]{IEEEtran} 
 \usepackage{amsmath,amssymb}
 \usepackage{subfigure}
 \usepackage{graphicx,graphics,color,psfrag}
 \usepackage{cite,balance}
 \usepackage{algorithm}
 \usepackage{accents}
 \usepackage{amsthm}
 \usepackage{bm}
 \usepackage{url}
 \usepackage{algorithmic}
 \usepackage[english]{babel}
 \usepackage{multirow}
 \usepackage{enumerate}
 \usepackage{cases}
 \usepackage{stfloats}
 \usepackage{dsfont}
 \usepackage{color,soul}
 \usepackage{amsfonts}
  \usepackage{tcolorbox}

 \usepackage{cite,graphicx,amsmath,amssymb}
 \usepackage{subfigure}
 \usepackage{fancyhdr}
 \usepackage{hhline}
 \usepackage{graphicx,graphics}
 \usepackage{array,color}
 \usepackage{amsmath}
 \usepackage{amsthm}

\newtheorem{proposition}{Proposition}
\newtheorem{remark}{Remark}
\newtheorem{theorem}{Theorem}
\newtheorem{lemma}{Lemma}

\include{header}

\IEEEoverridecommandlockouts

\begin{document}

\title{Cooperative Interference Management for Over-the-Air Computation Networks}
\author{Xiaowen Cao, Guangxu Zhu, Jie Xu, and Kaibin Huang \\
\thanks{X. Cao is with the Future Network of Intelligence Institute (FNii), The Chinese University of Hong Kong (Shenzhen), Shenzhen, China, and the School of Information Engineering, Guangdong University of Technology, Guangzhou, China (e-mail: caoxwen@outlook.com).}
\thanks{G. Zhu is with Shenzhen Research Institute of Big Data, Shenzhen, China  (e-mail: gxzhu@sribd.cn). }
\thanks{J. Xu is with the Future Network of Intelligence Institute (FNii) and the School of Science and Engineering (SSE), The Chinese University of Hong Kong (Shenzhen), Shenzhen, China (e-mail: xujie@cuhk.edu.cn).  J.~Xu is the corresponding author.}
\thanks{K. Huang is with the Dept. of Electrical and Electronic Engineering, The University of Hong Kong, Pok Fu Lam, Hong Kong (e-mail: huangkb@eee.hku.hk). }
}

\markboth{}{}
\maketitle

\setlength\abovedisplayskip{2pt}
\setlength\belowdisplayskip{2pt}

\vspace{-1.5cm}
 \begin{abstract}\vspace{-0.4cm}
Recently, {\it over-the-air computation} (AirComp) has emerged as an efficient solution for {\it access points} (APs) to aggregate distributed data from many edge devices (e.g., sensors) by exploiting the waveform superposition property of multiple access (uplink) channels. While prior work focuses on the single-cell setting where inter-cell interference is absent,  this paper considers a multi-cell AirComp network limited by such interference and investigates the optimal policies for controlling devices' transmit power to minimize the {\it mean squared errors} (MSEs) in aggregated signals received at different APs.
First, we consider the scenario of centralized multi-cell power control. To quantify the fundamental AirComp performance tradeoff among different cells, we characterize the Pareto boundary of the multi-cell MSE region by minimizing the sum MSE subject to a set of constraints on individual MSEs.
Though the sum-MSE minimization problem is non-convex and its direct solution intractable, we show that this problem can be optimally solved via equivalently solving a sequence of convex {\it second-order cone program} (SOCP) feasibility problems together with a bisection search. This results in an efficient algorithm for computing the optimal centralized multi-cell power control, which optimally balances the interference-and-noise-induced errors and the signal misalignment errors unique for AirComp.
Next, we consider the other scenario of distributed power control, e.g., when there lacks a centralized controller. In this scenario, we introduce a set of {\it interference temperature} (IT) constraints, each of which constrains the maximum total inter-cell interference power between a specific pair of cells. Accordingly, each AP only needs to individually control the power of its associated devices for single-cell MSE minimization, but subject to a set of IT constraints on their interference to neighboring cells.
By optimizing the IT levels, the distributed power control is shown to provide an alternative method for characterizing the same multi-cell MSE Pareto boundary as the centralized counterpart. Building on this result, we further propose an efficient algorithm for different APs to cooperate in iteratively updating the IT levels to achieve a Pareto-optimal MSE tuple, by pairwise information exchange.
Last, simulation results demonstrate that cooperative power control using the proposed algorithms can substantially reduce the sum MSE of AirComp networks compared with the conventional single-cell approaches.
\end{abstract}
\vspace{-0.4cm}
\begin{IEEEkeywords}\vspace{-0.3cm}
Over-the-air computation, multi-cell cooperation, power control, interference management, interference temperature.
\end{IEEEkeywords}
\vspace{-0.3cm}
\section{Introduction}\label{sec:intro}
One common operation of future {\it Internet-of-Things} (IoT) is to aggregate sensing data or computation results transmitted by many edge devices (e.g., sensors and smart phones).
Recently, {\it over-the-air computation} (AirComp) has emerged as a promising solution for such fast {\it wireless data aggregation} (WDA) as required by ultra-low-latency and high-mobility applications  \cite{nomo_function_Nazer,Gastpar08,nomo_function_Abari}.
The core idea of AirComp is to exploit the signal-superposition property of a {\it multiple access channel} (MAC) for ``over-the-air aggregation". This enables an {\it access point} (AP) to directly receive the aggregated version of the simultaneously transmitted data from devices. The sharing of the whole spectrum by all devices overcomes the issue of long latency faced in massive access. With proper pre-processing at devices and post-processing at AP, AirComp can go ahead averaging to compute a class of so-called nomographic functions (e.g., geometric mean and polynomial functions).
As a result, AirComp finds a wide range of applications ranging from distributed sensing \cite{Gastpar08,nomo_function_Abari} to distributed consensus \cite{MolinariConsensus} to distributed machine learning \cite{GX18_learning,GX20,Kang19,Tao_19,Deniz20}. The theme of this paper is to design techniques for cooperative interference management to facilitate the large-scale implementation of AirComp in multi-cell networks.

\vspace{-0.3cm}
\subsection{Over-the-Air Computation }
The concept of AirComp was first studied from the information-theoretic perspective in \cite{nomo_function_Nazer}, where structured codes are designed to exploit interference arising from simultaneous transmissions for fast functional computation over a MAC. Subsequently, a strong result was proved in \cite{Gastpar08} that simple AirComp with uncoded analog transmission is optimal in terms of minimizing the noise-induced distortion in WDA, which we term AirComp error, if the sensing data sources are independent and identically Gaussian distributed \cite{Gastpar08}.
Another vein of research on AirComp focuses on the signal processing perspective  \cite{Xiao08,Wang11,Abari15}. The optimal scheme for power allocation that minimizes the AirComp error was studied targeting distributed signal estimation from noisy observations \cite{Xiao08}. In \cite{Wang11}, the authors proposed power allocation schemes under a different criterion of minimum outage probability where an outage event occurs when the AirComp error exceeds a certain threshold. One requirement for implementing AirComp is the synchronization between transmissions by devices.
A solution for meeting the requirement was proposed in \cite{Abari15}, in which the AP broadcasts a reference-clock signal to all devices.

Most recent advancements in AirComp have led to its integration with more complex wireless techniques and systems, and its new application to the area of distributed machine learning. {\it Multiple-input-multiple-output} (MIMO) AirComp was developed to exploit spatial multiplexing for supporting vector-valued functional computation targeting multi-modal sensing \cite{GX18,DZ18}.
The channel feedback overhead in MIMO AirComp was then exempted in \cite{ZD18_blind}, by solving a bilinear estimation problem that can recover both the channel information and the desired functions simultaneously from a set of noisy received aggregated signals. In the fast growing area of distributed machine learning, AirComp finds a new application in efficiently enabling an edge server to aggregate distributed learning results transmitted by edge devices \cite{GX18_learning,GX20,Kang19,Tao_19,Deniz20}.

Power control for AirComp, which is the theme of this work, concerns controlling the transmit power of energy-constrained edge devices to cope with channel fading and noise that can potentially result in unacceptable AirComp errors. The simple transmission scheme of channel inversion is widely adopted in the AirComp literature to overcome fading so that multiuser signals arriving at an AP are aligned in magnitude, which is required for receiving a desired functional value of distributed data \cite{GX18,DZ18,GX18_learning,Kang19}.
However, it is well established that the channel inversion incurs severe noise amplification when channels are in deep fade, resulting in large AirComp errors.
To address this issue, the corresponding power control needs to be jointly designed over devices with the objective of minimizing the AirComp error.
This differs from the power control in conventional systems with different objectives of enhancing data rates or ensuring link reliability, which result in well-known policies such as water filling or channel truncation (see, e.g., \cite{Goldsmith}).
Based on the metric of {\it mean squared error} (MSE), the optimal power control policies were studied for AirComp under individual power constraints in \cite{Cao_fading,Tao_19,wanchun19} and under a sum power constraint in \cite{Xiao08,Wang11}, which were found to have different structures from their counterparts for conventional wireless communication systems. For instance, to minimize MSE, devices with relatively weak links should transmit with full power but others should perform channel inversion \cite{Cao_fading}, achieving a balance between the noise-induced errors and the signal misalignment errors.

While the prior work assumes a single-cell network, we envision the large-scale deployment of AirComp in a multi-cell network, to support ubiquitous coverage for next-generation IoT. This leads to simultaneous AirComp tasks in different cells, each of which is characterized by its application and corresponding data type (e.g., sensing or learning) as well as aggregation function (e.g., averaging or geometric mean). The coexistence of different AirComp tasks, however, affects each other due to the inter-cell interference. This gives rise to the new challenge of managing such interference by multi-cell cooperation so as to rein in the errors in the coexisting tasks.

\vspace{-0.3cm}
\subsection{Cooperative Interference Management}
Cooperative interference management for conventional radio access networks is a well-studied area (see, e.g., \cite{Gesbert2010} and the references therein). A wide range of relevant techniques and issues have been studied such as  beamforming \cite{Pareto,TomLuo}, network throughput \cite{LiuLiang2012,Etkin2008} and power control \cite{Kandukuri2002,JianweiHuang2009}. However, the challenges faced by designing cooperative interference management for AirComp networks differ from those for conventional radio access networks as they provide different services. A conventional radio access network is designed to support radio access to users. In contrast, the function of an AirComp network is to perform WDA over devices that are either sensors or workers. This results in different operations and performance metrics for the conventional radio access networks and the emerging AirComp networks. In terms of operations, the former suppresses multiuser interference so as to support multiuser data streams while the latter aggregates simultaneous data streams to compute a desired function. In terms of performance metrics, the conventional ones measure rates or reliability (e.g., sum rate \cite{Pareto,LiuLiang2012} and outage probability \cite{Kandukuri2002}) while those for AirComp should measure the accuracy in the received functional value (e.g., MSE). Cooperative interference management for AirComp networks remains a largely uncharted area. Recently, an initial study in this area was reported in \cite{Lan2020}, where a scheme called simultaneous signal-and-interference alignment is proposed to maximize the number of interference-free aggregated data streams in a two-cell AirComp network. In this work, we study the same theme of multi-cell cooperation but from a different perspective, namely power control.

%
\vspace{-0.3cm}
\subsection{Main Contributions}

In this paper, we consider an AirComp network comprising multiple cells. In each cell, one AP serves as a fusion center to aggregate date from multiple devices in the same cell. The aggregated signal received at each AP is exposed to inter-cell interference due to simultaneous uplink transmissions by devices in the neighboring cells. A novel framework of coordinated power control for managing such interference to suppress the errors in coexisting AirComp tasks is presented in the current work. The main contributions of this work are summarized as follows.

\begin{itemize}
	\item { \bf Multi-cell MSE tradeoffs with centralized  coordinated power control:}
	In this scenario, the transmit power of all devices is subject to centralized control by a centralized network controller. First, to understand the fundamental MSE performance trade-offs among these cells, we  characterize the Pareto boundary of the AirComp MSE region of simultaneous AirComp tasks in different cells using the so-called {\it MSE-profiling} technique. This is equivalent to minimizing the sum MSE of all APs subject to a set of MSE constraints for individual cells and individual transmit power constraints at devices, in which the devices' transmit powers and APs' signal scaling factors for noise suppression, called {\it denoising factors}, are jointly optimized.
	The problem is non-convex due to the coupling between power control variables and denoising factors.
	Though the direct solution is intractable, we propose an alternative approach to obtain its optimal solution by equivalently solving a sequence of problems, each being a convex {\it second order cone program} (SOCP), combined with a simple bisection search. This leads to an efficient algorithm of computing the optimal policy for centralized coordinated power control, which optimally balances the suppression of the interference-and-noise-induced errors and signal misalignment errors.
	\item {  \bf Distributed power control with interference-temperature coordination:}
	We consider another scenario of distributed power control, where the centralized controller is unavailable.
The distributed power control is realized by introducing a set of {\it interference temperature} (IT) constraints, each of which limits the maximum power of total interference from one cell to the other.
Given the IT constraints, the multi-cell power control reduces to single-cell operations, where each AP only needs to control the power of its associated devices for single-cell MSE minimization. While power control is distributed, multi-cell cooperation is realized by optimizing the IT levels.
It is shown that by proper IT levels control, the same MSE Pareto boundary as the centralized counterpart can be achieved.
To materialize the gain promised by such optimality, we further propose an efficient algorithm for different APs to cooperate in iteratively updating the IT levels for practically achieving a Pareto-optimal MSE tuple, by only pairwise information exchange.
Based on the algorithm, all cells are ensured to monotonically reduce their individual MSE values from the starting point corresponding to no cooperation, providing incentives for the cells to cooperative.
   \item  {\bf Performance evaluation:} Simulation results are presented to validate the derived analytical results. It is shown that both the centralized and distributed implementation of coordinated power control can substantially improve the AirComp performance, compared with the conventional design
       without cooperation.
\end{itemize}

The remainder of the paper is organized as follows.
Section \ref{sec:system} presents the system model of the AirComp networks.
Sections \ref{sec_cen} and \ref{sec_dis} present the centralized and distributed power control for characterizing the Pareto boundary of MSE region, respectively.
Finally, Section \ref{sec_simu} presents the simulation results, followed by the conclusion in Section \ref{sec_con}.

{\it Notations:}  Bold lowercase and uppercase letters refer to column vectors and matrices, respectively.
$\mathbb{E}(\cdot)$ denotes the expectation operation, and the superscript $T$ represents the transpose operation.
For a complex number $a$, ${\rm Re}\{a\}$ denotes the real part and the superscript ${\dagger}$ denotes the conjugate operation.
For a vector $\bm a$, $\|\bm a\|_2$ denote the Euclidean norm. $|\bm A|$ denotes the determinant of a squared matrix $\bm A$.
\section{System Model and Performance Metrics}\label{sec:system}

\subsection{System Model}
\begin{figure}
\centering
 \setlength{\abovecaptionskip}{-4mm}
\setlength{\belowcaptionskip}{-4mm}
    \includegraphics[width=5in]{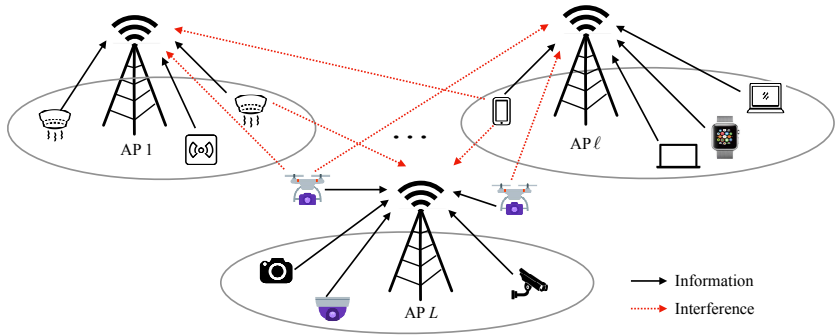}
\caption{A multi-cell AirComp network, where, in each cell, an AP aims at  aggregating data from its associated edge devices. } \label{fig:model}
\vspace{-0.5cm}
\end{figure}
We consider a multi-cell AirComp network with multiple APs as shown in Fig. \ref{fig:model}, where each AP acting as a fusion center aggregates sensing data (e.g., temperature, humidity) from edge devices.
In each cell, the aggregated signal received at the AP is exposed to the inter-cell interference caused by uplink transmission of devices in neighboring cells.
Let ${\mathcal L}\triangleq\{1,2,\cdots,L\}$ denote the set of $L$ APs, each dealing with a heterogeneous type of interested data, and ${\mathcal K_{\ell}}\triangleq\{\sum_{i=1}^{\ell-1}K_i+1,\cdots,\sum_{i=1}^{\ell-1}K_i+K_{\ell}\}$ denote the set of $K_{\ell}\ge 0$ edge devices collecting sensing readings associated with AP $\ell\in\cal L$ with $K_0\triangleq 0$ and ${\mathcal K_{\ell}}\cap {\mathcal K_j}=\emptyset, \forall {\ell}\neq j,\ell,j\in\cal L$.
Let ${\mathcal K}$ denote the set of all $K$ devices with ${\mathcal K}\triangleq {\mathcal K_1}\cup {\mathcal K_2}\cup\cdots\cup {\mathcal K_{L}}$ and $K=\sum\limits_{\ell\in\mathcal L} K_{\ell}$.
Specifically, AP $\ell$ needs to estimate the average of the type-$\ell$ data from the $K_{\ell}$ devices in $\mathcal K_l$.
Let $X_{k}$ denote the sensing reading measured by device $k\in\mathcal K_{\ell}$ associated with AP $\ell\in\cal L$, which is assumed to be {\it independent and identically distributed} (i.i.d.) over devices.
The desired average of type-$\ell$ data at AP $\ell$, denoted by $\tilde {f_{\ell}}(\cdot)$, is given by
\begin{align}
\tilde f_{\ell}= \frac{1}{K_{\ell}} \left(\sum \limits_{k\in{\mathcal K_{\ell}}}X_{k}\right),~\forall \ell\in\cal L.\label{desired_functions}
\end{align}
%
To facilitate power control, $X_k$ is normalized as $s_{k} \triangleq \Psi_{\ell}(X_{k} )$, $\forall k\in {\mathcal K_{\ell}}, \ell\in\cal L$ \cite{Cao_fading}. The linear function $\Psi_{\ell}(\cdot)$ denotes the normalization operation to ensure that $\{s_{k} \}_{k\in{\mathcal K_{\ell}}}$ have zero mean and unit variance, assuming $\{X_k\}_{k\in{\mathcal K_{\ell}}}$ have identical means and variance.
Upon receiving the average of transmitted data $\{s_{k} \}_{k\in{\mathcal K_{\ell}}}$ at each AP $\ell\in\cal L$, i.e.,
\begin{align}
f_{\ell} =\frac{1}{K_{\ell}} \sum \limits_{k\in{\mathcal K_{\ell}}}s_{k} , ~\forall \ell\in\cal L, \label{functions}
\end{align}
it can simply recover the desired $\tilde f_{\ell}  $ from $f_{\ell} $ via the de-normalization operation as follows:
\begin{align}
\tilde f_{\ell}   =\Psi_{\ell}^{-1}(f_{\ell}  ),\label{target_functions}
\end{align}
in which $\Psi_{\ell}^{-1}(\cdot)$ represents the inverse function of $\Psi_{\ell}(\cdot)$.
Therefore, with the one-to-one mapping between $f_{\ell} $ and $\tilde f_{\ell} $, we refer to $f_{\ell} $ as the target-function value in this paper.

To design adaptive power control, it is sufficient to consider a single realization of channels and analyze the control policy as a function of the channel states.
Let $h_{k}$ and $g_{k,j}, \forall k\in{\mathcal K_{\ell}} $ denote the channel coefficient of the data link between device $ k\in{\mathcal K_{\ell}}$ and its associated AP $\ell\in{\mathcal L}$, and that of the interference link between device $k\in \mathcal K_{\ell}$ and non-associated AP $j\in{\mathcal L}\setminus \{\ell\}$, respectively.
Let $b_{k} $ denote the transmit coefficient at device $k\in{\mathcal K_{\ell}}, \ell \in \cal L$ for transmitting information to AP $\ell\in\cal L$.
Therefore, the received signal at each AP $\ell$ is
\begin{align}\label{yl}
y_{\ell} = \sum \limits_{k\in{\mathcal K_{\ell}}}h_{k} b_{k} s_{k} \! + \sum \limits_{j\in{\mathcal L}\setminus \{{\ell}\}}\sum \limits_{i\in{\mathcal K_{j}}}\!  g_{i,\ell} b_{i} s_{i } +{ w}_{\ell} , \forall \ell\in{\mathcal L},
\end{align}
where $w_{\ell}  \sim  \mathcal{CN}(0, \sigma^2)$ models channel noise at the AP $\ell$.
To invert the data link for signal alignment at the AP, the transmit coefficient $b_{k} $ is set as $b_k=\frac{ \sqrt{p_{k} } h_{k}^{\dagger} }{ |h_{k} |}$, where $p_{k} \ge 0$ denotes the transmit power at device $k$ that is a control variable of our interest. 		
Then \eqref{yl} reduces to
\begin{align}
y_{\ell} = \sum \limits_{k\in{\mathcal K_{\ell}}}|h_{k} |\sqrt{p_{k} }s_{k} \! + \sum \limits_{j\in{\mathcal L}\setminus \{{\ell}\}} \sum \limits_{i\in{\mathcal K_{j}}}\! \tilde{g}_{i,\ell} \sqrt{p_{i} } s_{i} +{ w}_{\ell},~ \forall \ell\in{\mathcal L},\label{time_Yv}
\end{align}
where $\tilde{g}_{i,\ell}\triangleq \frac{ g_{i,\ell} h_{i}^{\dagger} }{ |h_{i} |}, i\in{\mathcal K_{j}}, j\in\mathcal{L}\setminus \{\ell\}$, represents the effective interference channel to AP $\ell$.
Following the existing approach (see, e.g. \cite{Cao_fading}), the signal $y_{\ell} $ is scaled at AP $\ell$ using a denoising factor denoted by $\eta_{\ell}$.
The scaled signal is given as
\begin{align}\label{ave_functions}
\hat f_{\ell}   =  \frac{{\rm Re}\{y_{\ell} \}}{{K_{\ell}}\sqrt{\eta_{\ell}}}.
\end{align}
Furthermore, in practice, each device $k\in\mathcal K_{\ell}$ is constrained by a maximum power budget $\bar P_{k}$:
\begin{align}
p_{k}  \leq \bar{P}_{k},~\forall k\in{\mathcal K_{\ell}}, ~ \ell\in\cal L.\label{fad_bar_P_ave}
\end{align}

\vspace{-0.5cm}
\subsection{Performance Metrics
}
 We are interested in minimizing the distortion of the recovered average of the transmitted data, {\it with respect to} (w.r.t.) the ground truth average $f_{\ell} ,\forall \ell\in\cal L$ at each AP $\ell$.	
The AirComp error in cell $\ell$ is measured by the corresponding instantaneous MSE defined as
\begin{align}
 \widetilde{\rm MSE}_{\ell} ( \{p_{k} \}_{k\in{\mathcal K}}, \eta_{\ell} )&=\mathbb{E}\left[(\hat{f}_{\ell} -f_{\ell})^2\right]=\frac{1}{K_{\ell}^2}\mathbb{E}\left[\left(\frac{{\rm Re}\{y_{\ell} \} }{\sqrt{\eta_{\ell}}} - \sum \limits_{k\in{\mathcal K_{\ell}}}s_{k} \right)^2 \right] \notag \\
&=\frac{1}{K_{\ell}^2} \left(\sum \limits_{k\in{\mathcal K_{\ell}}}\left(\frac{ \sqrt{p_{k} }|h_{k} |}{\sqrt{\eta_{\ell}}}-1\right)^2+\frac{\sigma^2+\sum \limits_{j\in{\mathcal L}\setminus \{{\ell}\}}\sum \limits_{i\in{\mathcal K_{j}} } p_{i} |\hat{g}_{i,\ell} |^2}{\eta_{\ell}}\right),\label{MSE_L_3}
\end{align}
where the expectation is over the distribution of the transmitted signals $\{s_{k} \}_{k\in{\mathcal K}}$ and $\hat{g}_{k,\ell}={\rm Re}\{\tilde{g}_{k,\ell}\}$.
For notational convenience, we use ${\rm MSE}_{\ell}( \{p_{k} \}_{k\in{\mathcal K}}, \eta_{\ell} )=K_{\ell}^2\widetilde{\rm MSE}_{\ell} ( \{p_{k} \}_{k\in{\mathcal K}}, \eta_{\ell} )$  in the sequel to represent the MSE without the constant term $\frac{1}{K_{\ell}^2}$ in \eqref{MSE_L_3}.
It is observed from \eqref{MSE_L_3} that the ``intra-cell interference" is exploited to enable functional computation, while the inter-cell interference interferes with the operation.

\begin{figure}
\centering
 \setlength{\abovecaptionskip}{-4mm}
\setlength{\belowcaptionskip}{-4mm}
    \includegraphics[width=3.5in]{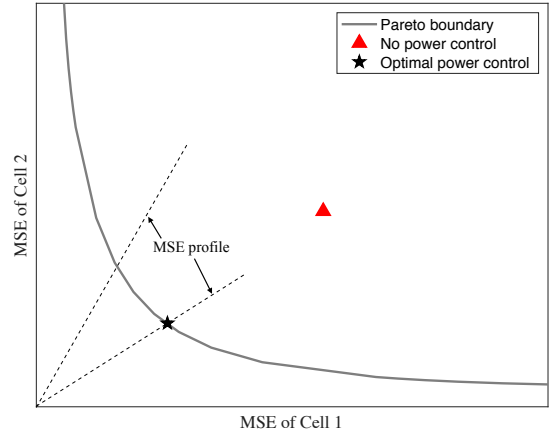}
\caption{Illustration of the Pareto boundary of MSE region based on the optimal coordinated power control. } \label{fig:Power_MSE_Region}
\vspace{-0.5cm}
\end{figure}

We define the {\it MSE region} for AirComp to be the set of MSE-tuples for all $L$ APs that can be simultaneously achievable for all $L$ APs under a given set of individual maximum power constraints for the devices, given as
\vspace{2mm}
\begin{equation}\label{Region}
\boxed{
\begin{aligned}
\bullet \;& \text{MSE Region}: \\
&{\mathcal M}\triangleq \bigcup_{\substack{ 0\leq p_{k}  \leq \bar{P}_{k},~\forall k\in{\mathcal K} ,\\
	  \eta_{\ell} \ge 0, ~\forall \ell\in\cal L}}\Big \lbrace \left(\Phi_1,\Phi_2,\cdots,\Phi_{L}\right):  \Phi_{\ell}\geq {\rm MSE}_{\ell} ( \{p_{k} \}_{k\in {\mathcal K}}, \eta_{\ell} ),  \forall \ell\in{\mathcal L} \Big \rbrace.
\end{aligned}
}\vspace{2mm}
\end{equation}
We are particularly interested in the operational points on the Pareto boundary (or, equivalently, the {\it Pareto optimal} points) of the MSE region $\mathcal M$, which corresponds to the lower-left boundary of this MSE region. Note that in the Pareto boundary, we can only reduce a particular AP's MSE at a cost of increasing the MSE at others, as illustrated in Fig. \ref{fig:Power_MSE_Region}.

\vspace{-0.5cm}
 \section{Centralized Power Control via MSE-Profiling}\label{sec_cen}


In this section, we focus on the scenario of centralized power control, when there exists a centralized controller with global {\it channel state information} (CSI) to coordinate all APs on the sum MSE reduction. In such a scenario,  we first introduce the so-called {\it MSE-profiling} technique to characterize the Pareto boundary of MSE region. Based on this, an optimal algorithm for joint power control and denoising factor design is presented to achieve the Pareto boundary.
\vspace{-0.3cm}
\subsection{Characterization of Pareto Boundary of MSE Region via MSE Profiling}\label{subsec_MSE}

Inspired by the ``rate profile" approach proposed in \cite{RateProfile}, which is a widely used approach to characterize the Pareto boundary of rate region in multiuser communication systems, we propose to characterize the Pareto boundary of the MSE region by using the MSE-profiling technique as presented in this subsection.
Define a particular MSE-profiling vector as ${\bm \beta} = [\beta_1,\beta_2,\cdots,\beta_{L}]$. Then the MSE-tuple on the Pareto boundary of the MSE region can be obtained by solving the following optimization problem with a specified MSE-profiling vector $\bm\beta$:
\begin{align}
\mathbf{(P1):}\min_{\{ p_{k} \}_{k\in{\mathcal K}},\{ {\eta}_{\ell} \}_{\ell\in\cal L}, \varepsilon\ge 0 }~& \varepsilon\notag\\
{\rm s.t.}~~~~~~~~&
 {\rm MSE}_{\ell}( \{p_{k} \}_{k\in{\mathcal K}}, \eta_{\ell} )\le\beta_{\ell} \varepsilon,~\forall \ell\in\cal L \label{P1_weighted_MSE} \\
&0\le p_{k}  \leq \bar{P}_{k},~\forall k\in{\mathcal K}\label{P2_P} \\
& \eta_{\ell} \ge 0,~\forall \ell \in {\mathcal L}, \label{P2_eta}
\end{align}
where $\varepsilon$ denotes the achievable sum MSE of the $L$ APs, and $\beta_{\ell}$ represents the target ratio of the $\ell$-th AP's achievable MSE to the sum MSE achieved by all the $L$ APs, $ \varepsilon$.
In general, we assume that $\beta_{\ell}\ge 0, \forall \ell\in\cal L$, and it holds $\sum \limits_{\ell\in\mathcal L} \beta_{\ell}=1$, in which a smaller $\beta_{\ell}$ means that AP $\ell$ has higher priority in minimizing the MSE, ${\rm MSE}_{\ell}( \{p_{k} \}_{k\in{\mathcal K}}, \eta_{\ell} )$.
With a given $\bm \beta$, we denote the optimal value of problem (P1) by ${\varepsilon}^{\rm  opt1}$.
Accordingly, the achieved MSE tuple ${\bm \beta}\varepsilon^{\rm  opt1}$ corresponds to the Pareto-optimal point, which is exactly the intersection point between a ray in the direction of $\bm\beta$ and the Pareto boundary of the MSE region as geometrically illustrated in Fig. \ref{fig:Power_MSE_Region}. Therefore, by varying the values of $\bm\beta$, solving problem (P1) can yield the complete Pareto boundary for the MSE region.

\vspace{-0.3cm}
\subsection{Algorithm for Centralized Power Control}

In this subsection, we present the algorithm to optimally  solve  problem (P1) with a given $\bm \beta$.
Note that problem (P1) is non-convex due to the coupling between power control $\{ p_{k} \}_{k\in{\mathcal K}}$ and denoising factors $\{ {\eta}_{\ell} \}_{\ell\in\cal L}$ in constraint \eqref{P1_weighted_MSE}, and thus is hard to solve directly.
Thereby, we first consider the optimization of denoising factors $\{\eta_{\ell}\}_{\ell\in\cal L}$ with any given $\{ p_{k} \}_{k\in{\mathcal K}}$, and then find the optimal  $\{ p_{k} \}_{k\in{\mathcal K}}$  to solve problem (P1) with the optimal $\{\eta_{\ell}\}_{\ell\in\cal L}$.

First, with any given $\{ p_{k} \}_{k\in{\mathcal K}}$, problem (P1) can be decoupled into $L$ subproblems each for optimizing $\eta_{\ell}$ to minimize the MSE at one AP $\ell$.
The $\ell$-th subproblem is written as
\begin{align}
\min_{{\eta}_{\ell}\ge0}~&\frac{1}{\beta_{\ell} } \left( \sum \limits_{k\in{\mathcal K_{\ell}}}\left(\frac{ \sqrt{p_{k} }|h_{k} |}{\sqrt{\eta_{\ell}}}-1\right)^2+\frac{\sigma^2+\sum \limits_{j\in{\mathcal L}\setminus \{{\ell}\}}\sum \limits_{i\in{\mathcal K_{j}} } p_{i} |\hat{g}_{i,\ell} |^2}{\eta_{\ell}}\right)\label{P1_MSE_eta}.
\end{align}
Let $\nu_{\ell}=1/\sqrt{{\eta}_{\ell}}$, then problem \eqref{P1_MSE_eta} can be transformed to a convex quadratic problem as
\begin{align}
\min_{{\nu}_{\ell}\ge0}~&\frac{1}{\beta_{\ell} } \left( \sum \limits_{k\in{\mathcal K_{\ell}}}\left(\sqrt{p_{k} }|h_{k} |\nu_{\ell} -1\right)^2+\left(\sigma^2+\sum \limits_{j\in{\mathcal L}\setminus \{{\ell}\}}\sum \limits_{i\in{\mathcal K_{j}} } p_{i} |\hat{g}_{i,\ell} |^2\right)\nu_{\ell}^2 \right)\label{P1_weighted_MSE_nu}.
\end{align}
By setting the first derivative of the objective function in problem \eqref{P1_weighted_MSE_nu} to be zero, we can obtain the optimal solution $\nu_{\ell}^*$ to problem \eqref{P1_weighted_MSE_nu}.
 As a result, the optimal solution to problem \eqref{P1_MSE_eta} is obtained as $\eta_{\ell}^*=(\frac{1}{\nu_{\ell}^*})^2,\forall \ell\in\cal L$, given in the following proposition.
\begin{proposition}\label{Lemma_fea1_n_gamma}\emph{
	With any given $\{ p_{k} \}_{k\in{\mathcal K}}$, the optimal ${\eta}_{\ell} $ to problem \eqref{P1_MSE_eta} is given by
\begin{align}\label{P1_eta}
	\eta_{\ell}^* =\left( \frac{ \sum \limits_{k\in{\mathcal K_{\ell}}} p_{k} |h_{k} |^2+\sum \limits_{j\in{\mathcal L}\setminus \{{\ell}\}}\sum \limits_{i\in{\mathcal K_{j}}} p_{i} |\hat{g}_{i,\ell} |^2 +\sigma^2}{ \sum \limits_{k\in{\mathcal K_{\ell}}} \sqrt{p_{k} }|h_{k} | } \right)^2,\forall \ell\in\cal L.
\end{align}
}
\end{proposition}
\begin{remark}[Interfence-and-Noise-Induced Error Reduction]\emph{
It is observed from \eqref{P1_eta} that the optimal $\eta_{\ell}^*$ is monotonically increasing w.r.t. the noise variance $\sigma^2$, the received power $\sum \limits_{k\in{\mathcal K_{\ell}}} p_{k} |h_{k} |^2$, and the interference power from other devices associated with other APs $\sum \limits_{j\in{\mathcal L}\setminus \{{\ell}\}}\sum \limits_{i\in{\mathcal K_{j}}} p_{i} |\hat{g}_{i,\ell} |^2$.
On the one hand, as $\sigma^2$  increases, a large denoising factor $\eta_{\ell}^*$ is needed for suppressing the dominant noise-induced error.
On the other hand, as the interference power increases, a relatively larger $\eta_{\ell}^*$ is in need for inter-cell interference suppression to enable reliable multi-cell AirComp.
 }
\end{remark}
Next, we optimize $\{ p_{k} \}_{k\in{\mathcal K}}$ and $\varepsilon$ by substituting $\eta_{\ell}^*$ in \eqref{P1_eta} into problem (P1). Thus, we have
\begin{align}\label{P1_pk_MSE}
\min_{\{ 0\le p_{k}  \leq \bar{P}_{k} \},\varepsilon\ge 0 }~&\varepsilon \\
{\rm s.t.}~~~~~&K_{\ell}-\frac{\left( \sum \limits_{k\in{\mathcal K_{\ell}}} \sqrt{p_{k} }|h_{k} |  \right)^2}{ \left( {\sum \limits_{k\in{\mathcal K_{\ell}}} p_{k} |h_{k} |^2+\sum \limits_{j\in{\mathcal L}\setminus \{{\ell}\}}\sum \limits_{i\in{\mathcal K_{j}}} p_{i} |\hat{g}_{i,\ell} |^2 +\sigma^2}\right) }\le\beta_{\ell}  \varepsilon ,~\forall \ell\in\cal L \notag.
\end{align}
In the following, we show that problem \eqref{P1_pk_MSE} can be optimally solved by equivalently solving a sequence of feasibility problems each for a fixed $\varepsilon$. Denoting $\psi_{\ell}=K_{\ell}-\beta_{\ell}\varepsilon$, we define a series of feasibility problems with given $\varepsilon$ as
\begin{align}
{\rm Find} ~& {\{ p_{k} \}} \label{fea1}\\
{\rm s.t.}~~&\psi_{\ell} \left( {\sum \limits_{k\in{\mathcal K_{\ell}}} p_{k} |h_{k} |^2+\sum \limits_{j\in{\mathcal L}\setminus \{{\ell}\}}\sum \limits_{i\in{\mathcal K_{j}}} p_{i} |\hat{g}_{i,\ell} |^2 +\sigma^2}\right)\le{\left( \sum \limits_{k\in{\mathcal K_{\ell}}} \sqrt{p_{k} }|h_{k} |  \right)^2} ,~\forall \ell\in\cal L \label{fea1_weighted_MSE}\\
& 0\le p_{k}\le{\bar P_{k}}, \forall k\in\cal K.\label{fea1_popwerbudget}
\end{align}
Recall that ${\varepsilon}^{\rm opt1}$ denotes the optimal value achieved by problem (P1).
 With any given sum-MSE target $ \varepsilon$, if problem \eqref{fea1} is feasible, then we have ${\varepsilon}^{\rm opt1}\leq \varepsilon$; otherwise, ${\varepsilon}^{\rm opt1}> \varepsilon$ holds. Therefore, we can solve problem  \eqref{P1_pk_MSE} by equivalently solving the feasibility problems in \eqref{fea1} with different $\varepsilon$ together with a bisection search over $\varepsilon$.

Therefore, it remains to solve problem \eqref{fea1} with given $\varepsilon$.
Notice that  $\psi_{\ell}$ must be nonnegative, as the MSE $\varepsilon$ is upper bounded by $K_{\ell}$.
Hence, the constraints in \eqref{fea1_weighted_MSE} can be re-written as
\begin{align}
\sqrt{\psi_{\ell} \left( {\sum \limits_{k\in{\mathcal K_{\ell}}} p_{k} |h_{k} |^2+\sum \limits_{j\in{\mathcal L}\setminus \{{\ell}\}}\sum \limits_{i\in{\mathcal K_{j}}} p_{i} |\hat{g}_{i,\ell} |^2 +\sigma^2}\right)}\le  \sum \limits_{k\in{\mathcal K_{\ell}}} \sqrt{p_{k} }|h_{k} | ,~\forall \ell\in\cal L \label{fea1_pk}.
\end{align}
By introducing auxiliary variables $q_{k}=\sqrt{p_{k} }$, $\forall k\in{\mathcal K_{\ell}}, \ell\in\cal L$, and letting ${\bm q}_{\ell}\triangleq\left[q_{K_{\ell-1}+1},\cdots,q_{K_{\ell}} \right]^T$, ${\bm h}_{\ell}=\left[|h_{K_{\ell-1}+1} |,\cdots,|h_{K_{\ell}}| \right]^T$, ${\bm g}_{j, \ell}=\left[|\hat{g}_{(K_{j -1}+1),\ell} |,\cdots,|\hat{g}_{K_{j}, \ell}| \right]$, ${\bm H}_{\ell}={\rm diag}({\bm h}_{\ell})^T$,
 and ${\bm G}_{j, \ell}={\rm diag}({\bm g}_{j, \ell} )$, $\forall \ell\in\mathcal L,~j\in{\mathcal L}\setminus \{\ell\}$, we can transform the constraints in \eqref{fea1_pk} or equivalently \eqref{fea1_weighted_MSE} into a set of {\it second order cone} (SOC) constraints as:
\begin{align}
	\sqrt{\psi_{\ell}}\| {\bm \Sigma}_{\ell} \|_2 \leq {\bm q}_l^T{\bm h}_{\ell},
	      ~ \forall \ell\in{\mathcal L}, \label{fea1_q}
\end{align}
where ${\bm \Sigma}_{\ell}=[ {\bm q}_{1}^T{\bm G}_{1,\ell},\cdots,{\bm q}_{\ell}^T{\bm H}_{\ell},\cdots,{\bm q}_{L}^T{\bm G}_{L,\ell},\sigma]^T$.
Then, problem \eqref{fea1} is reformulated as the following SOCP, which can be solved efficiently by convex optimization tools, e.g., CVX \cite{cvx}.
\begin{align}
{\rm Find} ~& \{ q_{k} \} \label{fea2_phi1}\\
{\rm s.t.}~	  &    0\le q_{k}\le \bar q_{k},~\forall k\in\cal K \\
& \eqref{fea1_q}, \notag
\end{align}
where ${\bar q_{k}}\triangleq \sqrt{\bar P_{k}}$, $\forall k\in{\mathcal K_{\ell}}, \ell\in\cal L$.
Denote $\{q_k^*\}_{k\in\cal K}$ as the optimal solution to problem \eqref{fea2_phi1} with any given sum-MSE target $\varepsilon$, then we have $p_k^*=(q_k^*)^2, \forall k\in\cal K$ as the optimal solution to problem \eqref{fea1}.
Based on the solution to problem \eqref{fea1} together with the bisection search over $\varepsilon$, the optimal $\varepsilon^{\rm opt1}$ to problem \eqref{P1_pk_MSE} is thus obtained. With the obtained ${\varepsilon}^{\rm opt1}$, we can accordingly attain the globally optimal power control $\{p_k^{\rm opt1}\}_{k\in\cal K}$ by solving problem \eqref{fea1}, as well as the global optimal denoising factor $\{ \eta_{\ell}^{\rm opt1}\} _{\ell\in\cal L}$ for problem (P1) based on Proposition~\ref{Lemma_fea1_n_gamma}.
In summary, the algorithm for optimally solving problem (P1) is presented in Algorithm 1.

\begin{table}[htp]
\begin{center}\vspace{0.1cm}
\hrule
\vspace{0.2cm} \textbf{Algorithm 1 for Optimally Solving Problem (P1)}\vspace{0.2cm}
\hrule \vspace{0.1cm} 
\begin{itemize}
\item[a)] {\bf Input}: Maximum power budgets $\{\bar{P}_{k}\}_{k\in\cal K}$, MSE-profiling vector ${\bm \beta}$.
\item[b)] {\bf Initialization}: Let $ \varepsilon^{\rm low}= 0$, $\varepsilon^{\rm high}=\min \limits_{l\in\mathcal L} \frac{K_{\ell}}{\beta_{\ell}}$.
\item[c)] {\bf Repeat}
    \begin{itemize}
    \item[1)]  Compute ${\varepsilon}=\frac{{\varepsilon}^{\rm low}+{\varepsilon}^{\rm high}}{2}$, and then solve problem \eqref{fea1} with given ${\varepsilon}$ and the optimal solution of $\{p_{k}\}_{k\in\cal K}$ being $\{p_k^*\}_{k\in\mathcal K}$.
    \item[2)]  If problem \eqref{fea1} is feasible, then set $ \varepsilon^{\rm high}= \varepsilon$;
            otherwise, set  $ \varepsilon^{\rm low}= \varepsilon$;
    \end{itemize}
\item[d)] {\bf Until} $|\varepsilon^{\rm high}-\varepsilon^{\rm low}|$ converges within a prescribed accuracy.
\item[e)] {\bf Set}  ${\varepsilon}^{\rm opt1}=\frac{{\varepsilon}^{\rm low}+{\varepsilon}^{\rm high}}{2}$ and $p_{k}^{\rm opt1}=p_{k}^*, \forall k\in{\mathcal K_{\ell}}, \ell\in\cal L$.
\item[f)] {\bf Compute} $\{\eta_{\ell}^{\rm opt1}\}$ based on \eqref{P1_eta} in Proposition~\ref{Lemma_fea1_n_gamma}.
\item[g)] {\bf Output}: Obtain the optimal solution $\{p_{k}^{\rm opt1}\}_{k\in\cal K}, \{\eta_{\ell}^{\rm opt1}\}_{\ell\in\cal L}$, and ${\varepsilon}^{\rm opt1}$ to problem (P1).
\end{itemize}
\hrule \vspace{0.1cm}
\end{center}\vspace{-1cm}
\end{table}

%
%

\vspace{-0.3cm}
\section{Distributed Power Control based on Interference Temperature}\label{sec_dis}
The optimal centralized power control algorithm in the previous section requires the full cooperation among all APs coordinated by a centralized controller, to achieve the Pareto boundary of the MSE region.
In this section, we consider a practical scenario where a centralized controller is unavailable and study the distributed power control by exploiting the IT technique.
It will be proved that the IT-based distributed power control can actually provide an alternative method for achieving the same Pareto boundary of the MSE region as the centralized counterpart.

\vspace{-0.5cm}
\subsection{Alternative Characterization of Pareto Boundary of MSE Region via IT control}
Different from the MSE-profiling-based design, in this subsection, we provide an alternative problem formulation based on the IT technique to characterize the Pareto Boundary of MSE region, which features (distributed) single-cell power control under the constraints of a set of ITs to limit its interference to the neighboring cells.
To this end, we first introduce a set of IT levels denoted by $\Gamma_{\ell ,j}$, which is the maximum interference power from all devices associated with AP $\ell$ to AP $j$, $\forall \ell\in\mathcal L,~j\in{\mathcal L}\setminus \{\ell\}$.
For the purpose of illustration, we denote $\bm \Gamma$ as an $L(L-1)\times 1$ vector composed of $\Gamma_{\ell ,j}$'s, and $\bm \Gamma_{\ell}$ as a $2(L-1)\times 1$ vector consisting of
 $\Gamma_{j, \ell}$'s and $\Gamma_{\ell ,j}$'s, $\forall j\neq \ell, j\in\cal L$, for any given $\ell \in\cal L$.
Accordingly, for each AP $\ell$, we replace the interference term $ \sum \limits_{i\in{\mathcal K_{j}} } p_{i} |\hat{g}_{i,\ell} |^2$ in the MSE formula $ {\rm MSE}_{\ell}( \{p_{k} \}_{k\in{\mathcal K}}, \eta_{\ell} )$ by $\Gamma_{j, \ell}$, and impose a set of IT constraints each for one neighboring AP $j, \forall j\neq \ell$. As a result, the MSE minimization is implemented at each AP $\ell\in \mathcal L$ individually, which is explicitly expressed as
 \begin{align}
\mathbf{(P2.1.\ell):}\min_{\{ p_{k} \}_{k\in\mathcal K_{\ell}}, {\eta}_{\ell}\ge0 } ~&  \sum \limits_{k\in{\mathcal K_{\ell}}}\left(\frac{ \sqrt{p_{k} }|h_{k} |}{\sqrt{\eta_{\ell}}}-1\right)^2+\frac{\sigma^2+\sum \limits_{j\in{\mathcal L}\setminus \{{\ell}\}}\Gamma_{j, \ell} }{\eta_{\ell}}\notag\\
{\rm s.t.}~~~~~& \sum \limits_{k\in{\mathcal K_{\ell}} } p_{k} |\hat{g}_{k,j} |^2\le \Gamma_{\ell ,j}  , \forall j\in{\mathcal L}\setminus \{\ell\} \label{SP_lT}\\
&0\le p_{k}  \leq \bar{P}_{k},~\forall k\in{\mathcal K_{\ell}}.
\end{align}
For notational convenience, we denote $\overline{\Phi_{\ell}} (\bm \Gamma_{\ell})$ as the optimal value of problem (P2.1.$\ell$) with any given $\bm \Gamma_{\ell}$, and denote $ \{ p_{k}^{\rm opt2}\}_{k\in{\mathcal K_{\ell}}}$ and $ \eta_{\ell}^{\rm opt2}$ as the optimal solution to problem (P2.1.$\ell$).

Before solving problem (P2.1.$\ell$), we show that via the IT control, the single-cell distributed power control in problem (P2.1.$\ell$) leads to a {\it parametric} characterization of the Pareto boundary of MSE region w.r.t. $\bm\Gamma$ in the following proposition.\vspace{-0.3cm}
\begin{proposition}[Pareto Optimality Based on Interference Temperature Control]\label{Lemma_IT_Pareto}\emph{
For any MSE-tuple $(\Phi_1, \cdots,\Phi_{L})$ on the Pareto boundary achieved by $\{ \tilde p_{k} \}_{k\in\mathcal K}$ and $ \{\tilde \eta_{\ell}\}_{\ell\in\cal L} $, there exist a set of corresponding IT levels $\bm \Gamma$, with $\Gamma_{\ell ,j} =\sum \limits_{k\in{\mathcal K_{\ell}} } \tilde p_{k} |\hat{g}_{k,j} |^2$, $\forall \ell\neq j,\ell, j\in\mathcal L$, such that $ \Phi_{\ell} =\overline{\Phi_{\ell}} ( \bm \Gamma_{\ell})$, $\forall \ell\in \mathcal L$, and $\{ \tilde p_{k} \}_{k\in\mathcal K_{\ell}}$ and $ {\tilde \eta}_{\ell} $ are the optimal solution to problem (P2.1.$\ell$) for each cell $\ell\in \cal L$ with given  $\bm \Gamma$.
		}
\end{proposition}\vspace{-0.5cm}
\begin{IEEEproof}
See Appendix~\ref{Proof_Lemma_IT_Pareto}.
\end{IEEEproof}
Based on Proposition \ref{Lemma_IT_Pareto},  it follows that by solving problem (P2.1.$\ell$) and exhausting $\bm\Gamma$, we can accordingly obtain the complete Pareto boundary of MSE region same as that have been characterized from problem (P1) via MSE profiling.
Notice that in the IT-based method, we need to determine $L(L-1)$ parameters in $\bm\Gamma$ in order to find each boundary point, while only $K$ parameters are needed for the MSE-profiling-based design in the previous section.
Nevertheless, the IT-based design enables distributed power control that is efficient for practical  implementation (as illustrated next), while the MSE-profiling-based design must be realized in a centralized manner.
Furthermore, in terms of CSI requirement, the IT-based design requires each AP to have access to only the CSI of channels in its own cell and the related IT constraints, without requiring the availability of the global CSI of the whole network at a centralized controller.

\vspace{-0.5cm}
\subsection{Distributed Power Control under IT Constraints}\vspace{-0.2cm}
In this subsection, the single-cell power control problem in (P2.1.$\ell$) is solved under the IT constraints.
Note that problem (P2.1.$\ell$) is non-convex w.r.t. $\{ p_{k} \}_{k\in\mathcal K_{\ell}} $ and ${\eta}_{\ell}$, due to the non-convex term $\frac{p_k}{\eta_{\ell}}$, and thus hard to solve optimally in general.
 To overcome the difficulty, we define $ \nu_{\ell}=1/\eta_{\ell}$ as the inverse of denoising factor, and introduce an auxiliary variable $Q_{k}=\sqrt{p_{k}\nu_{\ell} }$ for each device $k\in\mathcal K_{\ell}$, such that problem (P2.1.$\ell$) w.r.t. $\{ p_{k} \}_{k\in\mathcal K_{\ell}}$ and $\eta_{\ell}$ can be transformed into the following equivalent problem w.r.t. $\{ Q_{k} \}_{k\in\mathcal K_{\ell}}$ and $\nu_{\ell}$:
   \begin{align}
\mathbf{(P2.2.\ell):}\min_{\{ Q_{k} \ge 0 \}_{k\in\mathcal K_{\ell}}, \nu_{\ell} \ge  0} ~&  \sum \limits_{k\in{\mathcal K_{\ell}}} \left(|h_{k}|Q_{k}-1\right)^2+\nu_{\ell}\left(\sigma^2+\sum \limits_{j\in{\mathcal L}\setminus \{{\ell}\}}\Gamma_{j, \ell}\right) \notag\\
{\rm s.t.}~~~~&  \sum \limits_{k\in{\mathcal K_{\ell}} } Q_{k}^2 |\hat{g}_{k,j} |^2\le \Gamma_{\ell ,j}\nu_{\ell}  , \forall j\in{\mathcal L}\setminus \{\ell\}\label{dis_Q_nu_lt}\\
&Q_{k}^2 \leq \bar{P}_{k}\nu_{\ell},~\forall k\in{\mathcal K_{\ell}}.\label{dis_Q_nu_P}
\end{align}
It can be easily verified that problem (P2.2.$\ell$) is jointly convex w.r.t. $\{Q_k\}_{k\in\mathcal K_{\ell}}$ and $\nu_{\ell}$, and thus can be efficiently solved by the standard convex optimization methods such as the interior point method \cite{Boyd2004}.
Alternatively, for gaining more design insights, we use the Lagrange duality method \cite{Boyd2004} to obtain a well-structured optimal solution on power control. We denote $ \{ Q_{k}^{\rm opt2}\}_{k\in{\mathcal K_{\ell}}}$ and $ \nu_{\ell}^{\rm opt2}$ as the optimal solution to problem (P2.2.$\ell$).
Let $\lambda_{\ell ,j}\ge 0$ denote the dual variable associated with the $j$-th IT constraint in \eqref{dis_Q_nu_lt} for problem (P2.2.$\ell$).
Then the partial Lagrangian of problem (P2.2.$\ell$) is given by
\begin{align*}
&{\mathcal L_{\ell}}\left( \{ Q_{k} \}_{k\in\mathcal K_{\ell}}, \nu_{\ell},\{\lambda_{\ell ,j}\}_{j\in{\mathcal L}\setminus \{\ell\}}\right)\\
&\!=\!\!\sum \limits_{k\in{\mathcal K_{\ell}}}\!\!\left(|h_{k}|^2\!+\!\sum \limits_{j\in{\mathcal L}\setminus \{{\ell}\}}\!\lambda_{\ell ,j} |\hat{g}_{k,j} |^2\!\right)\!\!Q_{k}^2\!-\!2\!\sum \limits_{k\in{\mathcal K_{\ell}}}\!|h_{k}|Q_{k}\!+\!\nu_{\ell}\left(\sigma^2+\sum \limits_{j\in{\mathcal L}\setminus \{{\ell}\}}\left(\Gamma_{j, \ell}- \!\lambda_{\ell ,j}\Gamma_{\ell ,j} \right)\!\right)\!+K_{\ell}.
\end{align*}
Then the dual function of problem (P2.2.$\ell$)  is given by
\begin{align}\label{dis_Q_nu.dual_function}
g_{\ell}(\{\lambda_{\ell ,j}\}_{j\in{\mathcal L}\setminus \{\ell\}}) = \min_{ \{ Q_{k} \ge0\}, \nu_{\ell}\ge0} &~{\mathcal L_{\ell}}\left( \{ Q_{k} \}_{k\in\mathcal K_{\ell}}, \nu_{\ell},\{\lambda_{\ell ,j}\}_{j\in{\mathcal L}\setminus \{\ell\}}\right)\\
{\rm s.t.}~~~~&Q_{k}^2 \leq \bar{P}_{k}\nu_{\ell},~\forall k\in{\mathcal K_{\ell}}.\label{Q_nu}
 \end{align}
\begin{proposition}\label{Lemma_lambda}\emph{
In order for the dual function $g_{\ell}(\{\lambda_{\ell ,j}\}_{j\in{\mathcal L}\setminus \{\ell\}})$ to be lower bounded, it must hold that $\sigma^2+\sum \limits_{j\in{\mathcal L}\setminus \{{\ell}\}}\left(\Gamma_{j, \ell}- \lambda_{\ell ,j}  \Gamma_{\ell ,j} \right)\ge 0$.}
\end{proposition}\vspace{-0.3cm}
\begin{IEEEproof}
See Appendix~\ref{Proof_Lemma_lambda}.
\end{IEEEproof}
The corresponding dual problem of problem (P2.2.$\ell$) is thus given by
\begin{align}\label{dis_Q_nu_dual_prob}
\max_{\{\lambda_{\ell ,j}\ge0\}} ~&g_{\ell}(\{\lambda_{\ell ,j}\}_{j\in{\mathcal L}\setminus \{\ell\}}) \\
{\rm s.t.}~~&\sigma^2+\sum \limits_{j\in{\mathcal L}\setminus \{{\ell}\}}\left(\Gamma_{j, \ell}- \lambda_{\ell ,j}  \Gamma_{\ell ,j} \right)\ge 0.
\end{align}
%
Since problem (P2.2.$\ell$) is convex and satisfies the Slater's condition \cite{Boyd2004}, strong duality holds between problem (P2.2.$\ell$) and its dual problem \eqref{dis_Q_nu_dual_prob}. As a result, we can solve problem (P2.2.$\ell$) by equivalently solving problem \eqref{dis_Q_nu_dual_prob}. For notational connivence, we denote the optimal solution to the dual problem \eqref{dis_Q_nu_dual_prob} as $\{\lambda_{\ell ,j}^{\rm opt2}\}_{j\in{\mathcal L}\setminus \{\ell\}}$, and that to problem \eqref{dis_Q_nu.dual_function} as $ \{ Q_{k}^*\}_{k\in{\mathcal K_{\ell}}}$ and $ \nu_{\ell}^*$ with any given $\{\lambda_{\ell ,j}\}$.
 In the following, we first evaluate the dual function $g_{\ell}(\{\lambda_{\ell ,j}\}_{j\in{\mathcal L}\setminus \{\ell\}})$ with any given $\{\lambda_{\ell ,j}\}_{j\in{\mathcal L}\setminus \{\ell\}}$ by solving problem \eqref{dis_Q_nu.dual_function}, and then obtain the optimal dual variable $\{\lambda_{\ell ,j}^{\rm opt2}\}_{j\in{\mathcal L}\setminus \{\ell\}}$ to maximize $g_{\ell}(\{\lambda_{\ell ,j}\}_{j\in{\mathcal L}\setminus \{\ell\}})$.

\subsubsection{Derivation of Dual Function}\label{Dis_DualFunction}

To obtain the dual function, we need to solve problem \eqref{dis_Q_nu.dual_function} equivalently with any given $\{\lambda_{\ell ,j}\}_{j\in{\mathcal L}\setminus \{\ell\}}$.

\noindent\underline{\it a) Optimizing $ \{Q_{k}\}_{ k\in{\mathcal K_{\ell}}}$ with Given $\nu_{\ell}$:}\label{Dis_Dual_Q_k_Title}
First, with any given inverse denoising factor $\nu_{\ell}$, problem \eqref{dis_Q_nu.dual_function} can be decomposed into the following $K_{\ell}$ subproblems, each for optimizing $Q_{k}, k\in\mathcal{K_{\ell}}$ as
\begin{align}
\min_{0\le Q_{k} \leq\sqrt{\bar{P}_{k}\nu_{\ell}} } ~&  \left(|h_{k}|Q_{k}-1\right)^2+\sum \limits_{j\in{\mathcal L}\setminus \{{\ell}\}}\lambda_{\ell ,j} Q_{k}^2 |\hat{g}_{k,j} |^2.  \label{dis_Q_nu_sub_ki}
\end{align}
By taking the first derivative of the objective function in problem \eqref{dis_Q_nu_sub_ki}, the optimal solution $Q_{k}^*$ is obtained as 
\begin{align}\label{Q_kl_opt1}
	Q_{k}^*=\min \left[ \sqrt{\bar{P}_{k}\nu_{\ell}}, \frac{|h_{k}| }{ |h_{k}|^2+\sum \limits_{j\in{\mathcal L}\setminus \{{\ell}\}}\lambda_{\ell ,j} |\hat{g}_{k,j} |^2}\right].
\end{align}

\noindent\underline{\it b) Optimizing $\nu_{\ell}$ with Obtained Optimal $ \{Q^*_{k}\}_{ k\in{\mathcal K_{\ell}}}$:}\label{Dis_Dual_nu_title}
Next, we find the optimal inverse denoising factor $\nu_l$ to problem \eqref{dis_Q_nu.dual_function} by substituting back the optimized $\{Q_k^*\}_{ k\in{\mathcal K_{\ell}}}$ in \eqref{Q_kl_opt1}. Before proceeding and to facilitate the description, we define $B_k$ as the {\it policy indicator} at device $k\in\mathcal K_{\ell}$, given by
\begin{align}\label{B_k}
	B_k\triangleq \frac{\bar{P}_{k} \left(|h_{k}|^2+\sum \limits_{j\in{\mathcal L}\setminus \{{\ell}\}}\lambda_{\ell ,j} |\hat{g}_{k,j} |^2 \right)^2}{|h_{k}|^2 }, \forall k\in\mathcal K_{\ell},
\end{align}
and assume that $B_1\leq \cdots \leq B_k \leq \cdots \leq B_{K_{\ell}}$ without loss of generality.
Notice that the value of $B_k$ determines the adopted power control policy (full power transmission or regularized channel inversion) at each device $k\in \mathcal K_{\ell}$ as will be discussed later.

Substituting $Q_k^{*}$'s in \eqref{Q_kl_opt1} into problem \eqref{dis_Q_nu.dual_function}, we obtain the following optimization problem:
 \begin{align}\label{dis_K_eta}
\min_{\nu_{\ell}\geq 0} ~ & F_{\ell}(\nu_{\ell}) \triangleq \sum \limits_{k\in{\mathcal K_{\ell}}} \min\left[   \left(|h_{k}|^2+\sum \limits_{j\in{\mathcal L}\setminus \{{\ell}\}}\lambda_{\ell ,j} |\hat{g}_{k,j} |^2\right)\bar{P}_{k}\nu_{\ell},\frac{|h_{k}|^2 }{ |h_{k}|^2+\sum \limits_{j\in{\mathcal L}\setminus \{{\ell}\}}\lambda_{\ell ,j} |\hat{g}_{k,j} |^2} \right]+K_{\ell}\notag\\
&~~~~-\!2\!\!\sum \limits_{k\in{\mathcal K_{\ell}}} \!\min \!\!\left[\!|h_{k}|\sqrt{\bar{P}_{k}\nu_{\ell}}, \frac{|h_{k}|^2 }{ |h_{k}|^2+\sum \limits_{j\in{\mathcal L}\setminus \{{\ell}\}}\lambda_{\ell ,j} |\hat{g}_{k,j} |^2}  \!\right]\!\!+\!\nu_{\ell}\!\!\left(\!\!\sigma^2\!+\!\!\sum \limits_{j\in{\mathcal L}\setminus \{{\ell}\}}\!\!\!\!\left(\!\Gamma_{j, \ell}\!-\! \lambda_{\ell ,j}  \Gamma_{\ell ,j} \right)\!\right).
\end{align}
Let $\nu_{\ell}^*$ denote the globally optimal solution to problem \eqref{dis_K_eta}.
To solve problem \eqref{dis_K_eta}, we first need to remove the ``min'' operation in the objective function  to simplify the derivation by adopting a \emph{divide-and-conquer} approach. In particular, we divide  the feasible set of problem \eqref{dis_K_eta}, namely $\{\nu_{\ell}\geq 0\}$, into $K_{\ell}+1$ intervals, each given by
\begin{align}\label{interval_F_k}
	{\mathcal F}_{\ell,k}=\left\{\nu_{\ell} \mid B_k \leq \frac{1}{\nu_{\ell}} \leq B_{k+1}\right\}, ~\forall k\in \{0\}\cup \mathcal K_{\ell},
\end{align}
where $B_{0}\triangleq 0$ and $B_{K_{\ell}+1}\to\infty$ are defined for notational convenience.
Then, we have
\begin{align}\label{static_eta_spec_domian}
\{\nu_{\ell}\geq 0\}=\bigcup \limits_{k\in \{0\}\cup\mathcal K_{\ell}} {\mathcal F}_{\ell,k}.
\end{align}
Given \eqref{static_eta_spec_domian}, solving problem \eqref{dis_K_eta} is equivalent to first solving $K_{\ell}+1$ subproblems (each for one interval ${\mathcal F}_{\ell,k}$, given as follows, $\forall k\in \{0\} \cup {\mathcal K}_{\ell})$, and then comparing their optimal values to obtain the minimum one:
 \begin{align}\label{F_k}
\min_{ \nu_{\ell}\in{\mathcal F}_{\ell,k}}  ~F_{\ell,k}(\nu_{\ell}),
\end{align}
where
 \begin{align}
&F_{\ell,k}(\nu_l) =
\left(\sum_{i=1}^k   \left(|h_{i}|^2+\sum \limits_{j\in{\mathcal L}\setminus \{{\ell}\}}\lambda_{\ell ,j} |\hat{g}_{i,j} |^2\right)\bar{P}_{i}+\sigma^2+\sum \limits_{j\in{\mathcal L}\setminus \{{\ell}\}}\left(\Gamma_{j, \ell}- \lambda_{\ell ,j}  \Gamma_{\ell ,j} \right)\right) \nu_{\ell}+k\notag\\
&~~-2 \left(  \sum_{i=1}^k|h_{i}|\sqrt{\bar{P}_{i}}\right)\sqrt{\nu_{\ell}}
+\sum_{n=k+1}^{K_{\ell}}\left(1- \frac{|h_{n}|^2 }{ |h_{n}|^2+\sum \limits_{j\in{\mathcal L}\setminus \{{\ell}\}}\lambda_{\ell ,j} |\hat{g}_{n,j} |^2} \right),~\forall k\in\mathcal K_{\ell}\setminus\{K_{\ell}\},\label{F_K_A}\\
	&F_{\ell,0}(\nu_{\ell}) = \sum_{i=1}^{K_{\ell}}\left(1- \frac{|h_{i}|^2 }{ |h_{i}|^2+\sum \limits_{j\in{\mathcal L}\setminus \{{\ell}\}}\lambda_{\ell ,j} |\hat{g}_{i,j} |^2} \right)+\nu_{\ell}\left(\sigma^2+\sum \limits_{j\in{\mathcal L}\setminus \{{\ell}\}}\left(\Gamma_{j, \ell}- \lambda_{\ell ,j}  \Gamma_{\ell ,j} \right) \right),\label{F_K_B}\\
	&F_{\ell,K_{\ell}}(\nu_{\ell})\!=\!\!\left(\!\sum_{i=1}^{K_{\ell}} \!  \left(\!|h_{i}|^2\!\!+\!\!\!\sum \limits_{j\in{\mathcal L}\setminus \{{\ell}\}}\!\!\lambda_{\ell ,j} |\hat{g}_{i,j} |^2\!\!\right)\!\bar{P}_{i}\!+\!\sigma^2\!+\!\!\!\!\sum \limits_{j\in{\mathcal L}\setminus \{{\ell}\}}\!\!\!\!\left(\!\Gamma_{j, \ell}\!- \!\lambda_{\ell ,j}  \Gamma_{\ell ,j} \!\right)\!\!\right)\! \!\nu_{\ell}\!-2\! \left(\!  \sum_{i=1}^{K_{\ell}}|h_{i}|\sqrt{\bar{P}_{i}}\!\right)\!\sqrt{\nu_{\ell}}\!+K_{\ell}.\label{F_K_C}
\end{align}
Suppose that $\nu_{\ell,k}^*$ and $F_{\ell,k}(\nu_{\ell,k}^*)$ denote the optimal solution and optimal value to the $k$-th subproblem in \eqref{F_k}.
By comparing the optimal values $\{F_{\ell,k}(\nu_{\ell,k}^*)\}$, we can obtain the optimal solution to problem \eqref{dis_K_eta}, given in the following proposition.
\begin{proposition}\label{Dis_nu_opt_with_kstar}\emph{
	The optimal $\nu_{\ell}^{*}$ for problem \eqref{dis_K_eta} is obtained as
\begin{align}\label{dualproblem_nu_1}
	\nu_{\ell}^{*}=\nu_{\ell,k^*}^*=\left(\frac{ \sum_{i=1}^{k^*} |h_{i}|\sqrt{\bar{P}_{i}}  }{ \sum_{i=1}^{k^*} \left(|h_i|^2+\sum \limits_{j\in{\mathcal L}\setminus \{{\ell}\}}\lambda_{\ell ,j} |\hat{g}_{i,j} |^2\right)\bar{P}_{i} +\sigma^2+\sum \limits_{j\in{\mathcal L}\setminus \{{\ell}\}}\left(\Gamma_{j, \ell}- \lambda_{\ell ,j}  \Gamma_{\ell ,j} \right) }\right)^2,
	\end{align}
	where $k^*=\arg \min_{k\in\mathcal K_{\ell}} F_{\ell,k}(\nu_{\ell,k}^*).$
	}
\end{proposition}\vspace{-1mm}
\begin{IEEEproof}
Please refer to Appendix~\ref{Dis_nu_opt_with_kstar_proof}.
\end{IEEEproof}
With $\nu_{\ell}^*$ at hands, $ \{ Q_{k}^*\}_{k\in{\mathcal K_{\ell}}}$ is derived accordingly.
Therefore, the optimal solution to problem \eqref{dis_Q_nu.dual_function} and also the dual function are obtained.

\subsubsection{Obtaining Optimal Dual Variables to Maximize Dual Function}
Next, it remains to search $\{\lambda_{\ell ,j}^{\rm opt2}\}_{j\in{\mathcal L}\setminus \{\ell\}}$ to maximize  $g_{\ell}(\{\lambda_{\ell ,j}\}_{j\in{\mathcal L}\setminus \{\ell\}})$ for solving dual problem \eqref{dis_Q_nu.dual_function}.
Since the dual function $g_{\ell}(\{\lambda_{\ell ,j}\}_{j\in{\mathcal L}\setminus \{\ell\}})$ is concave but non-differentiable in general, one can use subgradient based methods such as the ellipsoid method \cite{ellipsoid}, to obtain the optimal  $ \{\lambda_{\ell ,j}^{\rm opt2}\}$ for dual problem \eqref{dis_Q_nu_dual_prob}.
For the objective function in \eqref{dis_Q_nu.dual_function}, the subgradient w.r.t. $ \lambda_{\ell ,j}$ is $\sum \limits_{k\in{\mathcal K_{\ell}} } Q_{k}^2 |\hat{g}_{k,j} |^2- \Gamma_{\ell ,j}\nu_{\ell}$, $\forall j\in{\mathcal L}\setminus \{\ell\}$.

\subsubsection{Optimal Solution to Problems (P2.2.$\ell$) and (P2.1.$\ell$)}
With obtained $ \{\lambda_{\ell ,j}^{\rm opt2}\}_{j\in{\mathcal L}\setminus \{\ell\}}$, we can obtain the optimal solutions to problem (P2.2.$\ell$), i.e., $\{ Q_{k}^{\rm opt2}\}_{k\in{\mathcal K_l}}$ and $ \nu_{\ell}^{\rm opt2}$, according to \eqref{Q_kl_opt1} and \eqref{dualproblem_nu_1}, respectively.
After obtaining the optimal solution to problem (P2.2.$\ell$), the optimal solutions of $\{p_{k}^{\rm opt2}\}_{k\in\mathcal K_{\ell}}$ and $\eta_{\ell}^{\rm opt2}$ to problem (P2.1.$\ell$) can be correspondingly found by calculating $p_{k}^{\rm opt2}=\frac{(Q_{k}^{\rm opt2})^2}{\nu_{\ell}^{\rm opt2}}$ and $\eta_{\ell}^{\rm opt2}=\frac{1}{\nu_{\ell}^{\rm opt2}}$, as summarized in Theorem~\ref{eta_K_static_All}, for which the proof is omitted for brevity.
\begin{theorem} \label{eta_K_static_All}\emph{The optimal power control solution to problem (P2.1.$\ell$) is given by
	\begin{align}
		p_{k}^{\rm opt2}=\begin{cases}
		\bar{P}_{k}, ~&k\in\{1,\cdots,k^{\rm opt2} \},\\
			\frac{|h_{k}|^2\eta_{\ell}^{\rm opt2} }{\left( |h_{k}|^2+\sum \limits_{j\in{\mathcal L}\setminus \{{\ell}\}}\lambda_{\ell ,j}^{\rm opt2} |\hat{g}_{k,j} |^2\right)^2}, ~&k\in\{k^{\rm opt2}+1,\cdots,K_{\ell} \},			
		\end{cases}
	\end{align}
	where the threshold is given as 	
\begin{align}\label{dualproblem_nu}
\eta_{\ell}^{\rm opt2}=\left(\frac{ \sum_{i=1}^{k^{\rm opt2}} \left(|h_i|^2+\sum \limits_{j\in{\mathcal L}\setminus \{{\ell}\}}\lambda_{\ell ,j}^{\rm opt2} |\hat{g}_{i,j} |^2\right)\bar{P}_{i} +\sigma^2+\sum \limits_{j\in{\mathcal L}\setminus \{{\ell}\}}\left(\Gamma_{j, \ell}- \lambda_{\ell ,j}^{\rm opt2}  \Gamma_{\ell ,j} \right) }{ \sum_{i=1}^{k^{\rm opt2}} |h_{i}|\sqrt{\bar{P}_{i}}  }\right)^2,
\end{align}
with $k^{\rm opt2}=\arg \min \limits_{k\in\mathcal K_{\ell}} F_{\ell,k}(\nu_{\ell,k}^{\rm opt2})$.
}
\end{theorem}

\begin{remark}\emph{
From Theorem~\ref{eta_K_static_All}, it is observed that the optimal power control at each AP, derived from the MSE minimization problem under a set of IT constraints, exhibits a {\it threshold-based} structure. In particular, if the policy indicator of each device $k$ (i.e., a function of the power budget and channel quality of both the direct and interfering links given as $B_k$ in \eqref{B_k}) exceeds an optimized threshold (i.e., $B_k \geq\eta_{\ell}^{\rm opt2}$), then device $k$ will transmit with regularized channel inversion, where the regularization can balance the tradeoff between the signal-magnitude alignment and interference-induced error suppression; otherwise, device $k$ will employ the full power transmission.
Similar observations were also made in \cite{Cao_fading} studying single-cell AirComp. However, the policy indicator (also called quality indicator) defined in \cite{Cao_fading} is determined only by the power budget and the channel quality of the direct link, while that in the current work also accounts for the interference to other APs.
}	
\end{remark}

\vspace{-0.5cm}
\subsection{Efficient Algorithm for Optimizing IT Levels}\label{Updating_IT_Decentralized}
The optimization of IT levels provides a mechanism for harnessing the cooperation gain on top of the distributed power control. In this subsection, we present an efficient algorithm for updating $\bm \Gamma$ in an iterative manner with only peer-to-peer signaling between APs.
In each iteration, a particular pair of two APs updates their IT levels, which needs to ensure that  the achievable MSE values at both APs are reduced or at least not increased, and the MSE values at other APs are not affected.
To successfully implement such a design, suppose that there is a reliable backhaul link between each pair of APs, such that all different pairs of APs can communicate with each other to exchange their mutual IT levels.

Consider the update of the mutual IT levels for a particular pair of APs.
 Recall that Proposition \ref{Lemma_IT_Pareto} shows that for any MSE tuple on the Pareto boundary, there must exist a $\bm \Gamma$ such that the optimal solution to the corresponding problem (P2.1.$\ell$)'s  can lead to that MSE tuple. However, for any given $\bm \Gamma$, it is still unsure whether it can achieve a Pareto-optimal MSE tuple of AirComp networks.
Inspiring by the simple rule for updating the IT levels in conventional multi-cell communication networks (instead of AirComp) \cite{Pareto}, in the following proposition we present the necessary condition on the IT levels in order for the resultant MSE tuple to be Pareto optimal.

\begin{proposition}\label{Lemma_Con_Pareto}\emph{(Necessary Condition for Pareto Optimality)
With any given $\bm\Gamma$, if the optimal MSE values $\overline{\Phi_{\ell}} ( \bm\Gamma_{\ell})$'s in problem (P2.1.$\ell$) (or equivalently (P2.2.$\ell$)) are Pareto optimal, then for any pair of APs $\ell$ and $j$, it must hold that $|\bm D_{\ell ,j}|=0$, where $\bm D_{\ell ,j}$ is a $2\times 2$ matrix defined as
\begin{align}\label{bm_D}
	\bm D_{\ell ,j}=\left[
	      \begin{matrix}
	      	{\frac{\partial \overline{\Phi_{\ell}} (  \bm\Gamma_{\ell}) }{\partial \Gamma_{\ell ,j}  }}&{\frac{\partial \overline{\Phi_{\ell}} ( \bm\Gamma_{\ell}) }{\partial \Gamma_{j, \ell}  }}\\
	      	{\frac{\partial \overline{\Phi_j} ( \bm\Gamma_{j}) }{\partial
	      	\Gamma_{\ell ,j} }}&{\frac{\partial \overline{\Phi_j} ( \bm\Gamma_{j}) }{\partial \Gamma_{j, \ell}  }}
	      \end{matrix}
	      \right].
\end{align}
}
\end{proposition}
\begin{IEEEproof}
See Appendix~\ref{Proof_Lemma_Con_Pareto}.
\end{IEEEproof}
Notice that one can obtain each component of $\bm D_{\ell ,j}$ based on the primal and dual optimal solution to problem (P2.2.$\ell$) with any given $\bm\Gamma_{\ell}$ \cite{Pareto}.
To be specific, we have
\begin{align}\label{Partial_gamma_11}
	{\frac{\partial \overline{\Phi_{\ell}} ( \bm\Gamma_{\ell}) }{\partial \Gamma_{\ell ,j}  }}= -\lambda_{\ell ,j}^{\rm opt2}{\nu}_{\ell}^{\rm opt},
\end{align}
where $\{\lambda_{\ell ,j}^{\rm opt2}\}$ is the optimal dual variables associated with the constraints in \eqref{dis_Q_nu_lt} of problem (P2.2.$\ell$) and $ {\nu}_{\ell}^{\rm opt}$ is the optimal solution to problem (P2.2.$\ell$).
Furthermore,  we have
\begin{align}\label{Partial_gamma_12}
	{\frac{\partial \overline{\Phi_{\ell}}( \bm\Gamma_{\ell}) }{\partial \Gamma_{j, \ell}  }}= {\nu}_{\ell}^{\rm opt},
	\end{align}
and ${\frac{\partial \overline{\Phi_j} ( \bm\Gamma_{j}) }{\partial \Gamma_{\ell ,j} }}$ and ${\frac{\partial \overline{\Phi_j} ( \bm\Gamma_{j}) }{\partial \Gamma_{j, \ell}  }}$ can be derived similarly.
\begin{remark}[Tightness of IT Constraints]
\emph{Combining Propositions \ref{Lemma_IT_Pareto} and \ref{Lemma_Con_Pareto}, it is observed that, for any particular $\bm \Gamma$ corresponding to a Pareto-optimal MSE tuple of the AirComp networks, it must hold that $\sum \limits_{k\in{\mathcal K_{\ell}} } p_{k}^{\rm opt2} |\hat{g}_{k,j} |^2=\Gamma_{\ell ,j} $, $\forall \ell\in\mathcal L, j\in{\mathcal L}\setminus \{\ell\}$. This says that all the IT constraints at APs must be tight in order to achieve the Pareto optimality.
}
\end{remark}

Next, we present an efficient rule to update the IT level in a pairwise manner inspired by Proposition \ref{Lemma_IT_Pareto} and \cite{Pareto}. Let $\bm \Gamma^{\prime}$ denote the updated $\bm \Gamma$, where all the elements in $\bm \Gamma$ remain unchanged except $[\Gamma_{\ell ,j},\Gamma_{j, \ell}]^T$ that is replaced by
\begin{align}\label{Gamma_update_rule}
	[\Gamma_{\ell ,j}^{\prime},\Gamma_{j, \ell}^{\prime}]^T=[\Gamma_{\ell ,j},\Gamma_{j, \ell}]^T+\delta_{\ell ,j}\cdot\bm d_{\ell ,j},
\end{align}
where $\delta_{\ell ,j}$ is a sufficiently small step size, and $\bm d_{\ell ,j}$ is a vector satisfying $\bm D_{\ell ,j}\bm d_{\ell ,j}<0$ (component-wise).
For notational conciseness, let $\bm D_{\ell ,j}=\left[
	      \begin{matrix}
	      	a & b\\
	      	c &d
	      \end{matrix}
	      \right]$, then one feasible $\bm d_{\ell ,j}$ is given as \cite{Pareto}
\begin{align}\label{bm_d}
\bm d_{\ell ,j}={\rm sign} \left (bc-ad\right)\cdot \left[ \alpha_{\ell ,j}d-b,a-\alpha_{\ell ,j}c \right]^T,
\end{align}
where ${\rm sign}(x)=1$ if $x\ge 0$ and ${\rm sign}(x)=- 1$ otherwise; and $\alpha_{\ell ,j}\ge0$ is a control parameter determining the ratio between the MSE decrements for APs $\ell$ and $j$.

\begin{remark}\label{Remark_ratio_optimallity}
\emph{Provided a sufficiently small step size $\delta_{\ell ,j}$, we can set the control variable $\alpha_{\ell ,j}\ge 1$ (or $\alpha_{\ell ,j}\ge 1$) to ensure that a larger (smaller) MSE decrement is achieved by AP $\ell$ than that for AP $j$. Therefore, via adjusting $\alpha_{\ell ,j}$ from zero to infinity, we can achieve different points on the Pareto boundary with lower MSE at both APs $\ell$ and $j$ than that at the starting point (e.g., under the design without cooperation).
}
\end{remark}

In summary,  the detailed procedure for the distributed IT levels update is described as follows.
\begin{itemize}
\setlength{\itemsep}{0pt}
\setlength{\parsep}{0pt}
\setlength{\parskip}{0pt}
\setlength{\parindent}{0pt}
  \item []{\bf Step 1)}: APs $\ell$ and $j$, $\ell\neq j$, exchange the current IT levels through the backhaul link;
    \item []{\bf Step 2)}: APs $\ell$ and $j$ solve problems (P2.2.$\ell$) and (P2.2.$j$) individually to obtain the optimal $\{Q_{k}\}_{k\in\mathcal K_{\ell}}$ and $\nu_{\ell}$ (and accordingly the optimal $\{p_{k}\}_{k\in\mathcal K_{\ell}}$ and $\eta_{\ell}$), as well as $\{Q_{k}\}_{k\in\mathcal K_{j}}$ and $\nu_{j}$ (and accordingly the optimal $\{p_{k}\}_{k\in\mathcal K_{j}}$ and $\eta_{j}$), respectively;
  \item []{\bf Step 3)}: According to \eqref{Partial_gamma_11} and \eqref{Partial_gamma_12},  APs $\ell$ and $j$, $\ell\neq j$, individually compute the elements $a$ and $b$, as well as $c$ and $d$ in $\bm D_{\ell ,j}$, respectively;
  \item []{\bf Step 4)}: The two APs share the computed results with each other to construct $\bm D_{\ell ,j}$ with \eqref{bm_D} and $\bm d_{\ell ,j}$ according to \eqref{bm_d};
      \item []{\bf Step 5)}: Both APs $\ell$ and $j$, $\ell\neq j$ can update their IT levels $\{\Gamma_{\ell ,j}^{\prime}\}$ according to \eqref{Gamma_update_rule};
\end{itemize}
The above procedure is repeated among different AP pairs until each element of the matrix $|\bm D_{\ell ,j}|, \forall {\ell}\neq j$, is less than a sufficiently small threshold $D_0$.
In summary, the compete algorithm for updating the IT levels in a decentralized manner is presented in Algorithm~2.
\begin{table}[htp]\label{table_Dis}
\begin{center}\vspace{0.1cm}
\hrule
\vspace{0.2cm} \textbf{Algorithm 2 for Updating the IT Levels in a Decentralized Way}\vspace{0.2cm}
\hrule \vspace{0.1cm} 
\begin{itemize}
\item[a)] {\bf Initialization}: Let $\Gamma_{\ell ,j}\ge0, \forall {\ell}\neq j, \ell,j\in\cal L$.
\item[b)] {\bf Repeat}
    \begin{itemize}
     \item[] For $\ell=1,2,\cdots,L$, and $ j=1,2,\cdots,L$, $j\neq \ell$
     \begin{itemize}
    \item[1)]  APs $\ell$ and $j$ exchange the current IT levels, i.e., $ \Gamma_{\ell,j}$ and $ \Gamma_{j,\ell}$;
     \item[2)] AP $\ell$ computes $a$ and $b$ in $\bm D_{\ell ,j}$ with $ \bm\Gamma_{\ell}$ according to \eqref{Partial_gamma_11} and \eqref{Partial_gamma_12}, respectively;
     \item[3)] Similarly, AP $j$ computes $c$ and $d$ in $\bm D_{\ell ,j}$ with $\bm\Gamma_{j}$;
     \item[4)] AP $\ell$ sends the results $a$ and $b$ to AP $j$ for constructing $\bm D_{\ell ,j}$, and similarly AP $j$ sends its results $c$ and $d$ to AP $\ell$;
    \item[5)]  APs $\ell$ and $j$ update $\bm\Gamma_{\ell}$ and $\bm\Gamma_{j}$ according to \eqref{Gamma_update_rule};
 \end{itemize}
 \item [ ] End For
    \end{itemize}
\item[c)] {\bf Until} $|\bm D_{\ell ,j}|$ is lower than a predetermined threshold $D_0$, i.e.,$|\bm D_{\ell ,j}|\le D_0, \forall {\ell}\neq j$.
\end{itemize}
\hrule 
\label{algorithm_P1}
\end{center}\vspace{-0.5cm}
\end{table}

\begin{remark}[Incentivized Multi-Cell Cooperation]\label{Remark_MSE_reduction}
\emph{
Essentially, the distributed algorithm reduces the overall MSE achieved by all APs in a pairwise manner. In other words, in each iteration, only a selected pair of APs updates their IT levels for their own MSE reduction without affecting the MSE performance of other APs.
 In this way, the generated solution can approach the Pareto boundary of the MSE region with the MSE reduced at all APs as compared to the starting point (e.g., without cooperation). Thereby this provides incentives for APs (even with self-interests) to participate in the cooperation for MSE reduction. }
 \end{remark}

\section{Simulation Results}\label{sec_simu}

In this section, we provide simulation results to show the MSE performance of the multi-cell AirComp networks.
In the simulation, we consider an AirComp network with two cells in Figs. \ref{fig:Cen_V_P_equal}, \ref{fig:Cen_V_K}, \ref{fig:Centralized_Pareto}, and \ref{fig:MSE_Converge_Two}, where the APs are located with a distance of $40$ {\it meters} (m) and the horizontal coordinates of them are fixed as $(0,0)$ and $(0,40{\rm m})$, respectively.
We also study the three-cell case in Figs. \ref{fig:Centralized_Pareto_Three} and \ref{fig:MSE_Converge_Three}, in which the third AP's horizontal coordinate is fixed as $(20{\rm m},40{\rm m})$.
All the devices are randomly located in a circle with its associated AP located at the center and a radius of $20$ m.
The direct and interference links follow Rayleigh fading channel models, specified by
$h_k=\theta_0\left( {\Theta}/{\Theta_0}\right)^{-\zeta}\bar{h}_k$ and $g_{k,j}=\theta_0\left( {\Theta}/{\Theta_0}\right)^{-\zeta}\bar{g}_{k,j}$, where $\bar{h}_k$'s and $\bar{g}_{k,j}$'s are modeled as i.i.d. {\it circularly symmetric complex Gaussian} (CSCG) random variables with zero mean and unit variance, $\theta_0=-60$~dB corresponds to the path loss at the reference distance of $\Theta_0=10$~m, $\Theta$ denotes the distance from the transmitter to the receiver, and $\zeta=3$ is the pathloss exponent.
Furthermore, we define the per-device power budget by $P$, and set $ \bar{P}_{k}=P$ for each device $ k\in{\mathcal K_{\ell}},  \ell\in\cal L$, and noise variance $ \sigma^2=-120$~dBm.
\vspace{-0.3cm}
 \subsection{Benchmarking Schemes}
 For performance comparison, we consider the following benchmark schemes without any cooperation required among  APs.
\begin{itemize}
	\item {\it Full power transmission:} The full power transmission is applied for all devices by setting $p_k=\bar P_k, \forall k\in\cal K$, in problem (P1). This scheme requires the most primitive control requiring no CSI collection and signaling overhead.
		\item \emph{Power control by ignoring interference:} Each AP optimizes the power allocation and denoising factor based on its own local CSI without any cooperation with other APs. Thus, the MSE is minimized independently at each AP ignoring the inter-cell interference. This leads to the single-cell AirComp power control problem for each cell $\ell\in\mathcal L$, presented as in \cite{Cao_fading}, given as
 \begin{align}
\min_{\{ 0\le p_{k}  \leq \bar{P}_{k}\}_{k\in\mathcal K_{\ell}}, {\eta}_{\ell}\ge0 } ~&  \sum \limits_{k\in{\mathcal K_{\ell}}}\left(\frac{ \sqrt{p_{k} }|h_{k} |}{\sqrt{\eta_{\ell}}}-1\right)^2+\frac{\sigma^2 }{\eta_{\ell}}.\label{Bench_2}
\end{align}
	\item \emph{Power control with maximum interference:} Each AP can access the local CSI and obtain the estimated interference based on an initial stage with full power transmission at all devices.
	However, the actual transmit power from the interference links is unknown. As a result, each AP minimizes the worst-case MSE by treating the maximum interference power $\sum \limits_{j\in{\mathcal L}\setminus \{{\ell}\}}\sum \limits_{k\in{\mathcal K_{j}} } {\bar P}_{k} |\hat{g}_{k,\ell} |^2$ as noise. This leads to the following problem of the same form as that in \eqref{Bench_2}, which can thus be similarly solved:
      \begin{align}
\min_{\{ 0\le p_{k}  \leq \bar{P}_{k}\}_{k\in\mathcal K_{\ell}}, {\eta}_{\ell}\ge0 } ~&  \sum \limits_{k\in{\mathcal K_{\ell}}}\left(\frac{ \sqrt{p_{k} }|h_{k} |}{\sqrt{\eta_{\ell}}}-1\right)^2+\frac{\sigma^2+\sum \limits_{j\in{\mathcal L}\setminus \{{\ell}\}}\sum \limits_{i\in{\mathcal K_{j}} } \bar{P}_{i} |\hat{g}_{i,\ell} |^2 }{\eta_{\ell}}.
\end{align}
\end{itemize}

\subsection{Performance Evaluation of the Proposed Cooperative Interference Management} 

\begin{figure}
\centering
 \setlength{\abovecaptionskip}{-4mm}
\setlength{\belowcaptionskip}{-4mm}
    \includegraphics[width=0.55\textwidth]{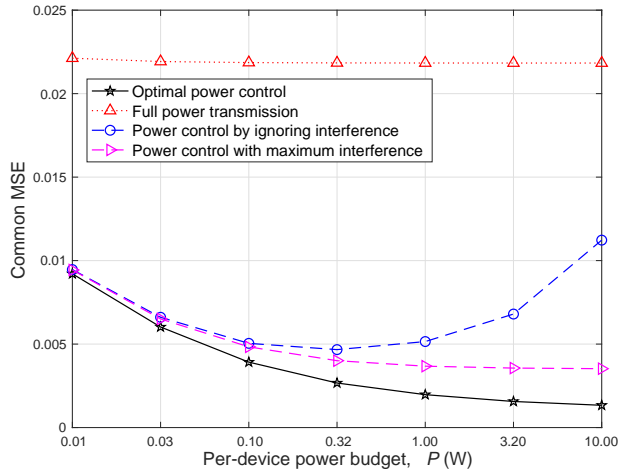}
\caption{Effect of per-device power budget on the MSE performance of multi-cell AirComp networks.} \label{fig:Cen_V_P_equal}
\end{figure}

We test the MSE performance by varying the per-device power budget in Fig.~\ref{fig:Cen_V_P_equal} with $L=2$ and $K_1=K_2=20$, where power budgets at different devices are set to be uniform. By setting $\beta_1=\beta_2=0.5$, the MSE metric is the common MSE between two cells.
Firstly, the proposed optimal centralized power control is observed to considerably outperform the other three benchmark schemes within the considered regimes.
At the low power budget regime (e.g., less than $0.1$ W), both the power control schemes by ignoring interference and with maximum interference can achieve close-to-optimal MSE performance, and all of the three schemes with power control  outperform the full-power-transmission scheme. This implies the effectiveness of power control optimization in suppressing the noise-induced error that is dominant for the MSE in the low  power budget regime.
As the power budget increases, the performance gap between the optimal centralized power control and the power-control-by-ignoring-interference becomes large, so as that between the power control schemes with maximum interference and by ignoring interference. This is due to the fact that the cooperative interference management is helpful in MSE reduction for AirComp networks, especially when the inter-cell interference becomes the main contributor for the MSE at the high power budget regime.
Besides, it is observed that the curves for both the optimal centralized power control and the power-control-with-maximum-interference schemes become saturated at the high power budget regime, meaning that the MSE performance is limited by the inter-cell interference and cannot be improved further by simply increasing the transmit power.

\begin{figure}
\centering
 \setlength{\abovecaptionskip}{-4mm}
\setlength{\belowcaptionskip}{-4mm}
    \includegraphics[width=0.55\textwidth]{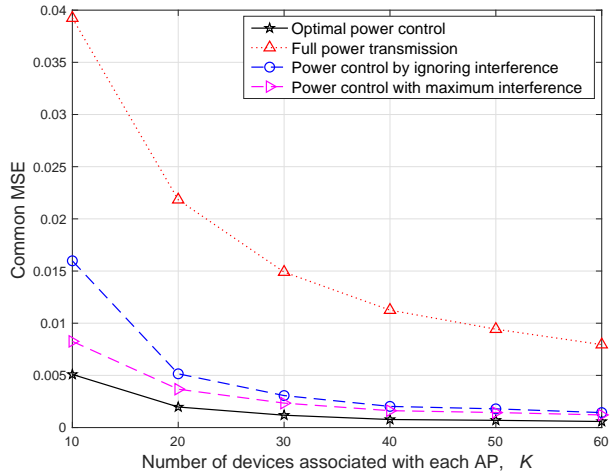}
\caption{Effect of the number of devices in each cell on the MSE performance of multi-cell AirComp networks.} \label{fig:Cen_V_K}
\vspace{-0.1cm}
\end{figure}

The effect of device population on the MSE performance is illustrated in Fig.~\ref{fig:Cen_V_K} with $\beta_1=\beta_2=0.5$ and $K_1=K_2=K$, where the power budgets at all devices are identically set to be $1$ W.
Firstly, it is observed that the MSE achieved by all the schemes decreases as $K$ increases, due to the fact that the AP receivers can aggregate more data for averaging.
Secondly, the performance gain achieved by the optimal power control over the benchmark schemes is significant throughout the whole regime of $K$, and especially prominent at the small $K$ regime.
The full-power-transmission scheme is observed to be the worst in MSE reduction among the others, showing the necessity of power control.
Furthermore, it is worth noting that, for all schemes with power control, the MSE performance is saturated at the large $K$ regime.
This is because that as $K$ increases, the inter-cell interference becomes the bottleneck of MSE reduction.

\begin{figure}[htbp]
  \centering
  \subfigure[Two-cell case.]
  {\label{fig:Centralized_Pareto}\includegraphics[width=8.1cm]{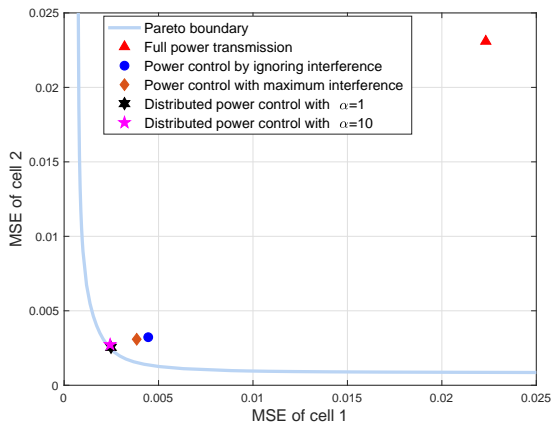}}
  \subfigure[Three-cell case.]
  {\label{fig:Centralized_Pareto_Three}
\includegraphics[width=8.1cm]{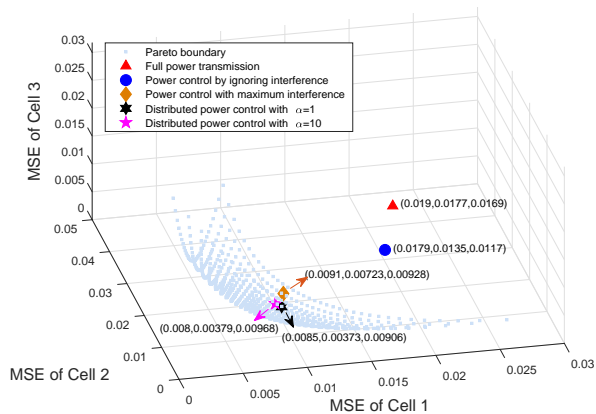}}
  \caption{MSE region of multi-cell AirComp networks.}
  \label{Fig:MSE_Pareto}
\vspace{-0.5cm}
\end{figure}

Figs.~\ref{fig:Centralized_Pareto} and \ref{fig:Centralized_Pareto_Three} show the MSE region of  AirComp networks with $L=2$ and $L=3$, respectively, where we set $P=1$ W, $K_{\ell}=20, \ell \in\mathcal L$, and $\alpha_{1,2}=\alpha_{1,3}=\alpha_{2,3}=\alpha$.
It is observed that all the benchmark schemes lie within the Pareto boundary that is obtained by the centralized power control by varying the MSE-profiling vector $\bm \beta$.
The distributed power control based on IT is observed to achieve the Pareto boundary.
Furthermore, through a comparison between the two cases in distributed power control with $\alpha=1$ and $\alpha=10$ under the two-cell AirComp network shown in Fig.~\ref{fig:Centralized_Pareto}, it is observed that a larger value of $\alpha$ bias the MSE minimization towards cell $1$ which is consistent with Remark \ref{Remark_ratio_optimallity}.
 Besides, the power-control-by-ignoring-interference scheme is observed to outperform the full-power-transmission scheme (closer to Pareto boundary), while the power-control-by-maximum-interference scheme outperforms the optimal-power-control-by-ignoring-interference scheme.
The former observation reveals the benefit of power control in minimizing the MSE performance of AirComp, while the later shows the effectiveness of the interference management.
Under the decentralized design, all the cells' MSE have reduced as compared to the case without cooperation, which is consistent with Remark \ref{Remark_MSE_reduction}.

\begin{figure}[htbp]
  \centering
  \subfigure[Two-cell case.]
  {\label{fig:MSE_Converge_Two}\includegraphics[width=8.1cm]{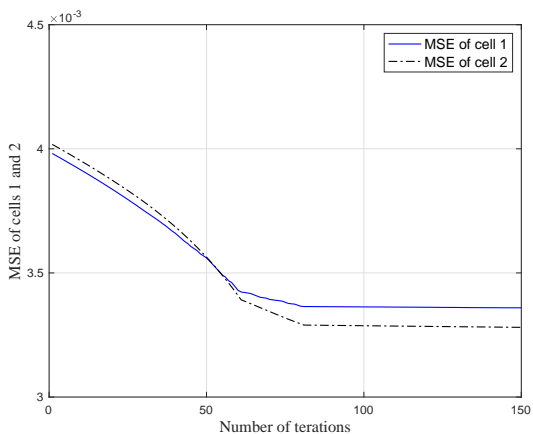}}
  \subfigure[Three-cell case.]
  {\label{fig:MSE_Converge_Three}
\includegraphics[width=8.1cm]{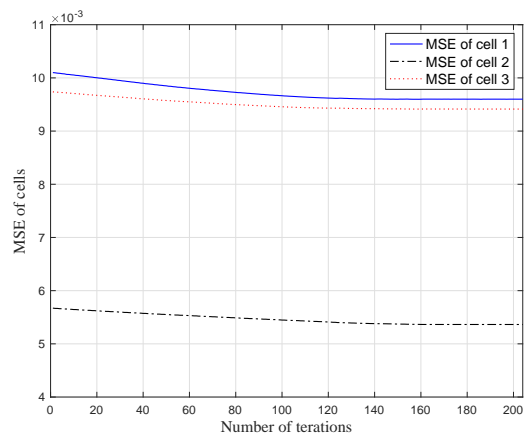}}
  \caption{Convergence analysis of the proposed distributed  IT-based power control.}
  \label{Fig:MSE_Converge}
\vspace{-0.5cm}
\end{figure}

Furthermore, the convergence performance of the pairwise decentralized algorithm in one channel realization is depicted for the setting of $\alpha =1$ and  $P=1$ W, in Figs.~\ref{fig:MSE_Converge_Two} and \ref{fig:MSE_Converge_Three}, corresponding to the two-cell and three-cell cases with each cell including $K_{\ell}=20$ devices, respectively.
The MSE values at all APs are monotonically non-increasing over time, thus validating Remark \ref{Remark_MSE_reduction} again in Section \ref{Updating_IT_Decentralized}, and showing the effectiveness of our design in practical decentralized implementation.

\section{Conclusion}\label{sec_con}
In this work, we considered multi-cell AirComp for which the power control was optimized over multiple cells to regulate the effect of inter-cell interference on AirComp performance.
Firstly, we considered the scenario of centralized multi-cell power control, based on which we characterized the Pareto boundary of the achievable multi-cell MSE region for AirComp networks. This is implemented by minimizing the overall MSE of all cells subject to a set of MSE-profiling constraints, which is solved via solving a sequence of convex SOCP feasibility problems together with a bisection search.
Next, we considered the scenario of distributed power control without a centralized controller, for which an alternative IT-based method was proposed to characterize the same MSE Pareto boundary, and enable a decentralized power control algorithm. In the decentralized design, each AP only needs to minimize its own MSE under a set of IT constraints, while different cells iteratively update the IT levels based on pairwise information exchange.
Remarkable performance gain in terms of AirComp accuracy was observed in the comparison with other designs without cooperation.

This work opens up several directions for further investigation on multi-cell AirComp. One direction is to develop multi-cell MIMO AirComp techniques for enabling coexisting vector-value function computation over multi-dimension data, where the key challenge lies in the joint design of multi-cell cooperative beamforming and power control.
Another interesting direction is to explore the cluster-based hierarchical design for large-scale AirComp, which needs to determine the optimal clustering policy for the interference suppression and computation distortion reduction.

%


\appendix
\subsection{ Proof of Proposition~\ref{Lemma_IT_Pareto}}\label{Proof_Lemma_IT_Pareto}
\vspace{-0.1cm}

Suppose that the given $\{ p_{k} \}_{k\in\mathcal K_{\ell}}$ and $ {\eta}_{\ell} $ can achieve the Pareto-optimal MSE tuple, and then we have the following MSE for each AP $\ell\in\cal L$:
\begin{align}\label{Phi_region}
	\Phi_{\ell}&=\sum \limits_{k\in{\mathcal K_{\ell}}}\left(\frac{ \sqrt{p_{k} }|h_{k} |}{\sqrt{\eta_{\ell}}}-1\right)^2+\frac{\sigma^2+\sum \limits_{j\in{\mathcal L}\setminus \{{\ell}\}}\sum \limits_{i\in{\mathcal K_{j}} } p_{i} |\hat{g}_{i,\ell} |^2}{\eta_{\ell}}, \forall \ell\in\cal L.
\end{align}
With  $\Gamma_{j, \ell} =\sum \limits_{i\in{\mathcal K_{j}} } p_{i} |\hat{g}_{i,\ell} |^2$, $\forall \ell\in\mathcal L, j\in{\mathcal L}\setminus \{\ell\}$, $\Phi_{\ell}$ in \eqref{Phi_region} can be rewritten as
\begin{align}\label{Phi_region_dis}
	\Phi_{\ell}&=\sum \limits_{k\in{\mathcal K_{\ell}}}\left(\frac{ \sqrt{p_{k} }|h_{k} |}{\sqrt{\eta_{\ell}}}-1\right)^2+\frac{\sigma^2+\sum \limits_{j\in{\mathcal L}\setminus \{{\ell}\}}\Gamma_{j, \ell}}{\eta_{\ell}}, \forall \ell\in\mathcal L.
\end{align}
Note that \eqref{Phi_region_dis} is derived similarly as the objective function of problem (P2.1.$\ell$) for each AP $\ell\in\cal L$. With $\Gamma_{\ell ,j} =\sum \limits_{k\in{\mathcal K_{\ell}} } p_{k} |\hat{g}_{k,j} |^2$, $\forall j\in{\mathcal L}\setminus \{\ell\}$, it is observed that $\{ p_{k} \}_{k\in\mathcal K_{\ell}}$ and $ {\eta}_{\ell} $ in \eqref{Region} satisfy all the constraints in problem (P2.1.$\ell$).
 Therefore, $\{ p_{k} \}_{k\in\mathcal K_{\ell}}$ and $ {\eta}_{\ell} $ must be the feasible solution for problem (P2.1.$\ell$).

 Hence, it remains to prove that $\{ p_{k} \}_{k\in\mathcal K_{\ell}}$ and $ {\eta}_{\ell} $ are exactly the optimal solution of problem (P2.1.$\ell$) for each AP $\ell$, where the achievable MSE accordingly is equal to the optimal value of problem (P2.1.$\ell$), i.e., $\Phi_{\ell}=\overline{\Phi_{\ell}} ( \bm \Gamma_{\ell})$.
 We prove this result by contradiction. Suppose that the optimal solution to problem (P2.1.$\ell$) in AP $\ell$ is denoted by $\{p_k^{\star}\}_{k\in\mathcal K_{\ell}}$ and $\eta_{\ell}^{\star}$ which are unequal to $\{ p_{k} \}_{k\in\mathcal K_{\ell}}$ and $ {\eta}_{\ell}$. Then, the new MSE achieved at AP $\ell$ denoted by $\phi_{\ell}$ is
\begin{align}
	\phi_{\ell}&=\sum \limits_{k\in{\mathcal K_{\ell}}}\left(\frac{ \sqrt{p_{k}^{\star} }|h_{k} |}{\sqrt{\eta_{\ell}^{\star} }}-1\right)^2+\frac{\sigma^2+\sum \limits_{j\in{\mathcal L}\setminus \{{\ell}\}}\Gamma_{j, \ell}}{\eta_{\ell}^{\star} }\notag\\
	&=\sum \limits_{k\in{\mathcal K_{\ell}}}\left(\frac{ \sqrt{p_{k}^{\star} }|h_{k} |}{\sqrt{\eta_{\ell}^{\star} }}-1\right)^2+\frac{\sigma^2+\sum \limits_{j\in{\mathcal L}\setminus \{{\ell}\}}\sum \limits_{i\in{\mathcal K_{j}} } p_{i} |\hat{g}_{i,\ell} |^2}{\eta_{\ell}^{\star} }<\Phi_{\ell}.
\end{align}
 As $\sum \limits_{k\in{\mathcal K_{\ell}} } p_{k}^{\star} |\hat{g}_{k,j} |^2\leq\Gamma_{\ell ,j}$, we further have the achievable MSE at any AP $j\neq \ell$ as
 \begin{align}
	\phi_j&=\sum \limits_{k\in{\mathcal K_j}}\left(\frac{ \sqrt{p_{k} }|h_{k} |}{\sqrt{\eta_j }}-1\right)^2+\frac{\sigma^2+\sum \limits_{i\in{\mathcal L}\setminus \{j,\ell\}}\sum \limits_{k\in{\mathcal K_{i}} } p_{k} |\hat{g}_{k,j} |^2+\sum \limits_{k\in{\mathcal K_{\ell}} } p_{k}^{\star} |\hat{g}_{k,j} |^2}{\eta_j }\notag\\
	&\leq\sum \limits_{k\in{\mathcal K_j}}\left(\frac{ \sqrt{p_{k} }|h_{k} |}{\sqrt{\eta_j }}-1\right)^2+\frac{\sigma^2+\sum \limits_{i\in{\mathcal L}\setminus \{j\}}\Gamma_{i,j}}{\eta_j }\notag\\
	&=\sum \limits_{k\in{\mathcal K_j}}\left(\frac{ \sqrt{p_{k}}|h_{k} |}{\sqrt{\eta_j }}-1\right)^2+\frac{\sigma^2+\sum \limits_{i\in{\mathcal L}\setminus \{j\}}\sum \limits_{k\in{\mathcal K_{i}} } p_{k} |\hat{g}_{k,j} |^2}{\eta_j }=\Phi_j.
\end{align}
Note that the MSE-tuple $(\phi_1,\cdots, \phi_{L})$ achieved by $\{\{p_k\}_{k\in\mathcal K_1},\eta_1\}$, $
\cdots$, $\{\{p_k^{\star}\}_{k\in\mathcal K_{\ell}},\eta_{\ell}^{\star}\}$ , $
\cdots$, $\{\{p_k\}_{k\in\mathcal K_{\ell}}, \eta_{\ell}\}$,  satisfies $\phi_{\ell}<\Phi_{\ell}$ and $\phi_j\leq \Phi_j$, $\forall j\neq \ell$, which contradicts the fact that $(\Phi_1,\cdots, \Phi_{L})$ is Pareto optimal. Thus the presumption cannot be true, and it holds that $\{ p_{k} \}_{k\in\mathcal K_{\ell}}$ and $ {\eta}_{\ell} $ are the optimal solution to problem (P2.1.$\ell$) for each AP $\ell$, i.e. $p_{k}=p_{k}^{\star}, \forall k\in\mathcal K_{\ell}$ and $\eta_{\ell}=\eta_{\ell}^{\star}$, and the achievable MSE is equal to the optimal value of problem (P2.1.$\ell$), i.e., $\Phi_{\ell}=\overline{\Phi_{\ell}} (\bm \Gamma_{\ell}), \forall \ell\in\cal L$.

\vspace{-0.3cm}
\subsection{ Proof of Proposition~\ref{Lemma_lambda}}\label{Proof_Lemma_lambda}
\vspace{-0.1cm}
Suppose $\sigma^2+\sum \limits_{j\in{\mathcal L}\setminus \{{\ell}\}}\left(\Gamma_{j, \ell}- \lambda_{\ell ,j}  \Gamma_{\ell ,j} \right)<0$. It thus follows that ${\mathcal L_{\ell}}\left( \{ Q_{k} \}_{k\in\mathcal K_{\ell}}, \nu_{\ell},\{\lambda_{\ell ,j}\}_{j\in{\mathcal L}\setminus \{\ell\}}\right)$ becomes negative infinity when $\nu_{\ell} \to +\infty$. This implies that the dual function $g_{\ell}(\{\lambda_{\ell ,j}\}_{j\in{\mathcal L}\setminus \{\ell\}})$ is unbounded from below in this case. Hence, it requires that $\sigma^2+\sum \limits_{j\in{\mathcal L}\setminus \{{\ell}\}}\left(\Gamma_{j, \ell}- \lambda_{\ell ,j}  \Gamma_{\ell ,j} \right)\ge 0$ to guarantee $g_{\ell}(\{\lambda_{\ell ,j}\}_{j\in{\mathcal L}\setminus \{\ell\}})$ to be bounded from below.

\vspace{-0.3cm}
\subsection{ Proof of Proposition~\ref{Dis_nu_opt_with_kstar}}\label{Dis_nu_opt_with_kstar_proof}
\vspace{-0.1cm}

To proceed with solving problem \eqref{dis_K_eta}, we alternatively solve each subproblem in \eqref{F_k} and then compare their optimal values $\{F_{\ell,k}(\nu_{\ell,k}^*)\}$ in \eqref{F_k}.
First, we have the following lemma.
\begin{lemma} \label{opt_noactive_lemma_k0}\emph{
The optimal solution $\nu_{\ell,0}^*$ to problem \eqref{F_k} when $ \nu_{\ell} \in{\mathcal F_{\ell,0}}$ is thus given by
\begin{align}\label{dis_nu_0}
	\nu_{\ell,0}^*&=\max \left[ \gamma_0 , ~\frac{1}{B_1}\right].
\end{align}
where $\gamma_0=\max \limits_{j\in{\mathcal L}\setminus \{{\ell}\}} \frac{1}{\Gamma_{\ell ,j}}\sum \limits_{k\in{\mathcal K_{\ell}}} \frac{|h_k|^2|\hat{g}_{k,j
	}|^2}{\left(|h_{k}|^2+\sum \limits_{j\in{\mathcal L}\setminus \{{\ell}\}}\lambda_{\ell ,j} |\hat{g}_{k,j} |^2\right)^2}$.
}
\end{lemma}\vspace{-0.3cm}
\begin{IEEEproof}
See Appendix~\ref{Proof_opt_noactive_lemma_k0}.
\end{IEEEproof}
Combining with IT constraints in \eqref{dis_Q_nu_lt}, it is worth to point out that if $\gamma_0 \geq \frac{1}{B_1}$, then the IT levels are unreasonably low, or the power budgets are unreasonably high, for which case all the power constraints become inactive.
This case does not make sense in practice and thus is excluded in the sequence discussion, and we hence assume the case with practical power budgets and the IT levels satisfying $\gamma_0 \leq \frac{1}{B_1}$.

Furthermore, for any $k\in \cal K$, the function $F_k(\nu_{\ell})$ is shown to be a unimodal function that first deceases in $[0, \tilde \nu_{k}]$ and then increases in $[\tilde \nu_{k}, \infty)$, where $\tilde \nu_{k}$ is the stationary point given by
\begin{align*}
	\tilde \nu_{k}&=\left(\frac{ \sum_{i=1}^{k} |h_{i}|\sqrt{\bar{P}_{i}}  }{ \sum_{i=1}^{k} \left(|h_i|^2+\sum \limits_{j\in{\mathcal L}\setminus \{{\ell}\}}\lambda_{\ell ,j} |\hat{g}_{i,j} |^2\right)\bar{P}_{i} +\sigma^2+\sum \limits_{j\in{\mathcal L}\setminus \{{\ell}\}}\left(\Gamma_{j, \ell}- \lambda_{\ell ,j}  \Gamma_{\ell ,j} \right) }\right)^2.
\end{align*}
Thus, the optimal solution $\nu_{\ell,k}^*$ to problem \eqref{F_k} when $ \nu_{\ell} \in\mathcal F_k$, $\forall k\in\mathcal K_{\ell}$ is thus given in the following lemma.
\vspace{-0.3cm}
\begin{lemma}\label{de_in_function_lemma}\emph{
The optimal solution $\nu_{\ell,k}^*$ to the $k$-th subproblem in \eqref{F_k} is given by
\begin{align}\label{static_eta_k}
	\nu_{\ell,k}^*=\max \left( \frac{1}{B_{k+1}}, ~\min\left( \tilde \nu_k,~ \frac{1}{B_{k}} \right)\right)=\tilde \nu_k.
\end{align}
	 }
\end{lemma}\vspace{-0.3cm}
\begin{IEEEproof}
Similarly as in \cite{Cao_fading}, it can be shown that
$1/B_k \le \tilde \nu_k \le 1/B_{k+1}$. This lemma thus follows directly.
\end{IEEEproof}
Therefore, with Lemma \ref{de_in_function_lemma} and by comparing the optimal values $\{F_{\ell,k}(\nu_{\ell,k}^*)\}$ among all subproblems, we can obtain the optimal solution to problem \eqref{dis_K_eta}.
This thus completes the proof.

\vspace{-0.5cm}
\subsection{ Proof of Proposition~\ref{Lemma_Con_Pareto}}\label{Proof_Lemma_Con_Pareto}

Based on Proposition~\ref{Lemma_IT_Pareto}, the optimal values of the problems in (P2.1.$\ell$) for all APs, $\{\overline{\Phi_{\ell}} (\bm \Gamma_{\ell})\}$, achieved by the optimal solution denoted by $\{p_{k}\}_{k\in\mathcal K_{\ell}}$ and $\{\eta_{\ell}\}_{\ell\in\cal L}$, correspond to a Pareto-optimal MSE-tuple, denoted by $(\Phi_1,\cdots, \Phi_{L})$, given as
\begin{align}
	\overline{\Phi_{\ell}} (\bm \Gamma_{\ell})=\Phi_{\ell}=\sum \limits_{k\in{\mathcal K_{\ell}}}\left(\frac{ \sqrt{p_{k} }|h_{k} |}{\sqrt{\eta_{\ell}}}-1\right)^2+\frac{\sigma^2+\sum \limits_{j\in{\mathcal L}\setminus \{{\ell}\}}\Gamma_{j, \ell}}{\eta_{\ell}}, \forall \ell\in\mathcal L.
\end{align}
Then we prove Proposition~\ref{Lemma_Con_Pareto} by contradiction.
Suppose that $|\bm D_{\ell ,j}|\neq 0$. With updated $\Gamma_{\ell ,j}^{\prime}$ and $\Gamma_{j, \ell}^{\prime}$ based on the updating rule of $\bm \Gamma^{\prime}$ in \eqref{Gamma_update_rule}, the optimal solutions to problems (P2.2.$\ell$) and (P2.2.$j$) for cells $\ell$ and $j$ are changed to be $\{\{p_{k}^{\star}\}_{k\in\mathcal K_{\ell}},\eta_{\ell}^{\star}\}$ and $\{\{p_{k}^{\star}\}_{k\in\mathcal K_j},\eta_j^{\star}\}$, respectively, while for those to problems (P2.2.$i$) for cell $i\neq \ell,j$, the optimal solutions remain unchanged.
Accordingly, the new achievable MSE for any AP $\forall i \neq \ell,j$, is given by
\begin{align}
	\phi_i&=\sum \limits_{k\in{\mathcal K_i}}\left(\frac{ \sqrt{p_{k}}|h_{k} |}{\sqrt{\eta_i}}-1\right)^2+\frac{\sigma^2+\sum \limits_{\bar\imath\in{\mathcal L}\setminus \{\ell,j,i\}}\sum \limits_{k\in{\mathcal K_{\bar\imath}} } p_{k} |\hat{g}_{k,i} |^2+\sum \limits_{k\in{\mathcal K_{\ell}} } p_{k}^{\star} |\hat{g}_{k,i} |^2+\sum \limits_{k\in{\mathcal K_{j}} } p_{k}^{\star} |\hat{g}_{k,i} |^2}{\eta_i}\notag\\
	&=\sum \limits_{k\in{\mathcal K_i}}\left(\frac{ \sqrt{p_{k}}|h_{k} |}{\sqrt{\eta_i}}-1\right)^2+\frac{\sigma^2+\sum \limits_{\bar\imath\in{\mathcal L}\setminus \{\ell,j,i\}}\Gamma_{\bar\imath, i}+\sum \limits_{k\in{\mathcal K_{\ell}} } p_{k}^{\star} |\hat{g}_{k,i} |^2+\sum \limits_{k\in{\mathcal K_{j}} } p_{k}^{\star} |\hat{g}_{k,i} |^2}{\eta_i}\le \Phi_i,\label{Lemma_update_Cell_i}
\end{align}
in which \eqref{Lemma_update_Cell_i} holds due to the fact that $\sum \limits_{k\in{\mathcal K_{\ell}} } p_{k}^{\star} |\hat{g}_{k,i} |^2\le \Gamma_{\ell ,i}$ and $\sum \limits_{k\in{\mathcal K_{j}} } p_{k}^{\star} |\hat{g}_{k,i} |^2\le \Gamma_{j,i}$.
 Then based on $\sum \limits_{k\in{\mathcal K_{j}} } p_{k}^{\star} |\hat{g}_{k,\ell} |^2\leq\Gamma_{j, \ell}^{\prime}$, we have the updated achievable MSE for AP $\ell$ as
\begin{align}
	\phi_{\ell}&=\sum \limits_{k\in{\mathcal K_{\ell}}}\left(\frac{ \sqrt{p_{k}^{\star} }|h_{k} |}{\sqrt{\eta_{\ell}^{\star} }}-1\right)^2+\frac{\sigma^2+\sum \limits_{n\in{\mathcal L}\setminus \{\ell,j\}}\sum \limits_{k\in{\mathcal K_{n}} } p_{k} |\hat{g}_{k,\ell} |^2+\sum \limits_{k\in{\mathcal K_{j}} } p_{k}^{\star} |\hat{g}_{k,\ell} |^2}{\eta_{\ell}^{\star} }\notag\\
	&=\sum \limits_{k\in{\mathcal K_{\ell}}}\left(\frac{ \sqrt{p_{k}^{\star} }|h_{k} |}{\sqrt{\eta_{\ell}^{\star} }}-1\right)^2+\frac{\sigma^2+\sum \limits_{n\in{\mathcal L}\setminus \{\ell,j\}}\Gamma_{n,\ell}+\sum \limits_{k\in{\mathcal K_{j}} } p_{k}^{\star} |\hat{g}_{k,\ell} |^2}{\eta_{\ell}^{\star} }\le\overline{\Phi_{\ell}}(  \bm \Gamma_{\ell}^{\prime} ).
\end{align}
Similarly, it also holds that $\phi_j\le\overline{\Phi_j}( \bm \Gamma_{j}^{\prime} )$ due to $\sum \limits_{k\in{\mathcal K_{\ell}} } p_{k}^{\star} |\hat{g}_{k,j} |^2\leq\Gamma_{\ell ,j}^{\prime}$.
Furthermore, based on \eqref{Gamma_update_rule} and $\bm D_{\ell ,j}\bm d_{\ell ,j}<0$, it follows that:
\begin{align*}
	\left[ \begin{matrix}
	  \phi_{\ell}\\
	     \phi_j
	      \end{matrix}\right]&\leq \left[ \begin{matrix}
	 \overline{\Phi_{\ell}}( \bm \Gamma_{\ell}^{\prime} )\\
	    \overline{\Phi_j}( \bm \Gamma_{j}^{\prime} )
	      \end{matrix}\right]
	     \cong \left[ \begin{matrix}
	 \overline{\Phi_{\ell}}(  \bm \Gamma_{\ell})\\
	    \overline{\Phi_j}( \bm \Gamma_{j})
	      \end{matrix}\right]+\delta_{\ell ,j}\cdot\bm D_{\ell ,j}\bm d_{\ell ,j}
	       < \left[ \begin{matrix}
	 \Phi_{\ell}\\
	     \Phi_j
	      \end{matrix}\right].
\end{align*}
Note that the achieved MSE-tuple $(\phi_1,\cdots, \phi_{L})$ based on the updated $ \bm \Gamma^{\prime} $, satisfies $\phi_{\ell}<\Phi_{\ell}$, $\phi_j\leq \Phi_j$, and $\phi_i\leq \Phi_i$, $\forall i\neq \ell,j$, which contradicts the fact that $(\Phi_1,\cdots, \Phi_{L})$ is Pareto-optimal. Thus the presumption is not true, and this proof is thus completed.

\vspace{-0.3cm}
\subsection{ Proof of Lemma~\ref{opt_noactive_lemma_k0}}\label{Proof_opt_noactive_lemma_k0}

Consider the case with $\nu_{\ell}\in{\cal F}_{\ell,0}$, for which $F_{\ell,0}(\nu_{\ell,0}) $ is linear and monotonically increasing w.r.t. $\nu_{\ell,0}$ based on Proposition~\ref{Lemma_lambda}.
In this case, we have $\nu_{\ell,0} \geq \frac{1}{B_1}$ accordingly.
Furthermore, with the IT constraint in \eqref{dis_Q_nu_lt}, there exists a possible solution in which only the IT constraint is tight instead of the power budget constraints. Thus, we correspondingly have the following potential constraints:
\begin{align}
	\nu_{\ell,0}\ge \sum \limits_{k\in{\mathcal K_{\ell}}} \frac{|h_k|^2|\hat{g}_{k,j}|^2}{\left(|h_{k}|^2+\sum \limits_{j\in{\mathcal L}\setminus \{{\ell}\}}\lambda_{\ell ,j} |\hat{g}_{k,j} |^2\right)^2 \Gamma_{\ell ,j}}, \forall j\in{\mathcal L}\setminus \{\ell\}.
\end{align}
Therefore, it is evident that the optimal solution to problem \eqref{F_k} when $ \nu_{\ell} \in{\mathcal F_{\ell,0}}$ can be given in Lemma~\ref{opt_noactive_lemma_k0}.
This completes the proof.

\end{document}